\documentclass[aps,prb,showpacs,twocolumn,amsmath,amssymb]{revtex4-2}

\usepackage{tabularx}
\usepackage{bm}
\usepackage{graphicx}

\usepackage{hyperref}
\hypersetup{colorlinks=true,urlcolor= blue,citecolor=blue,linkcolor= blue,bookmarks=true,bookmarksopen=false}

\usepackage{color}

\usepackage{amsmath,mathtools}
\usepackage{multirow}
\usepackage{dcolumn}
\usepackage{amssymb,amscd,xypic,bm,wasysym}
\usepackage{float}
\usepackage{cleveref}
\usepackage[caption=false,position=top,captionskip=0pt,farskip=0pt]{subfig}
\usepackage{soul}

\begin{document}
	
\title{Electronic chiralization as an indicator of the anomalous Hall effect in unconventional magnetic systems}
	
\author{Hua Chen}
\affiliation{Department of Physics, Colorado State University, Fort Collins, Colorado 80523, USA}
\affiliation{School of Advanced Materials Discovery, Colorado State University, Fort Collins, Colorado 80523, USA}

\begin{abstract}
The anomalous Hall effect (AHE) can appear in certain antiferromagnetic metals when it is allowed by symmetry. Since the net magnetization is usually small in such anomalous Hall antiferromagnets, it is useful to have other physical indicators of the AHE that have the same symmetry properties as the latter and can be conveniently measured and calculated. Here we propose such indicators named as electronic chiralization (EC), which are constructed using spatial gradients of spin and charge densities in general periodic crystals, and can potentially be measured directly by scattering experiments. Such constructions particularly reveal the important role of magnetic charge in the AHE in unconventional magnetic systems with vanishing net magnetization. Guided by the EC we give two examples of the AHE when magnetic charge is explicitly present: A minimum honeycomb model inspired by the magnetic-charge-ordered phase of kagome spin ice, and skew scattering of two-dimensional Dirac electrons by magnetic charge.
\end{abstract}

\maketitle

\section{Introduction}\label{sec:introduction}

The anomalous Hall effect (AHE) describes the transverse flow of charge currents driven by a longitudinal electric field in the absence of external magnetic fields \cite{Hall_1881,Smith_AHE_1921,Nagaosa_AHE_RMP_2010}. The mechanisms of the AHE in ferromagnets have been well understood by now \cite{Karplus_1954,Smit_1958,Berger_1970,Ye_AHE_1999,Nagaosa_AHE_RMP_2010,Xiao_RMP_2010}. In recent years particular interests have been devoted to the AHE appearing in certain antiferromagnets with vanishing net magnetization \cite{Tomizawa_2009,Chen_2014,Kubler_2014,Nakatsuji_2015,Nayak_Mn3Ge_2016,Zhou_Mn3XN_2019,Gurung_Mn3XN_2019,Boldrin_Mn3XN_2019,Zhao_Mn3NiN_2019,Liu_Mn3Pt_2018,Surgers_Mn5Si3_AHE_2014,Smejkal_CHE_2020,Li_QAHE_2019}, in contrast to the conventional wisdom that the anomalous Hall response is proportional to net magnetization. Although it is now clear that the AHE is generally nonvanishing as long as it is not forbidden by symmetry, it remains an open question whether one can find a convenient indicator of the AHE, that is similar to the net magnetization as a gauge-invariant observable but is not small because of energetic reasons in the AHE antiferromagnets. Such indicators, once identified, can help to understand the existence and variation of the AHE in inhomogeneous and disordered systems or across phase transitions, the dependence of the AHE on reorientation of the microscopic spin density field, and the scaling of the AHE with continuously tunable parameters such as temperature, doping, and strain, etc.

There have been a couple of proposals on constructing such indicators of the AHE in general magnetic crystals \cite{Suzuki_Cluster_2017,Hayami_AHE_AMD_2021}, based on the idea of multipole expansion. Ref.~\onlinecite{Suzuki_Cluster_2017} considered the magnetic (spherical) multipole moments of a finite atomic cluster having the same point group symmetry as the parent magnetic crystal. By decomposing the representation of a given point group in the basis of such cluster magnetic multipoles, the ones that resemble that formed by a magnetic dipole can be identified. The basis functions of such irreducible representations then transform in the same way as the magnetic dipole or the Hall conductivity vector under symmetry operations in the given point group, but not necessarily so under general O(3) operations applied on the whole magnetic crystal. More recently, Reference~\onlinecite{Hayami_AHE_AMD_2021} proposed to use the anisotropic magnetic dipole (AMD) as an indicator of the AHE. The AMD is a time-reversal-odd pseudovector and transforms in the same way as the Hall conductivity vector under general O(3) operations. However, to calculate the AMD for a given magnetic structure one still needs to first construct a finite atomic cluster using the approaches of \cite{Suzuki_Cluster_2017} or \cite{Suzuki_multipole_2019}.

The difficulty of defining multipole moments of an infinite crystal is long standing. In classical electromagnetism it is known that only the lowest-order nonvanishing multipole moment of a given charge or current distribution is independent of the choice of origin. Moreover, it has been realized through the studies on the electric polarization \cite{king-smith_1993,Resta_1994}, orbital magnetization \cite{xiao_2005,thonhauser_2005,shi_2007}, and magnetic toroidization \cite{Spaldin_2008,gao_2018} that even these low-order multipole moments cannot be directly obtained from the local charge and current densities in a unit cell, but require the information of the ground state wavefunctions of the whole crystal. However, for the purpose of finding a physical indicator of the AHE in antiferromagnets, it is more convenient to base the construction on readily available data of the magnetic structure such as that from neutron and X-ray experiments. Such scattering experiments directly probe the Fourier components of spin and charge distributions in a periodic crystal which themselves are gauge-invariant quantities. Moreover, constructions based on scattering amplitudes may allow direct determination of the AHE indicators without having to first fix the magnetic order. Finally, the indicators may point out new mechanisms or prototypical examples for the AHE in unconventional magnetic systems.

In this paper we propose a new class of indicators of the AHE, which we name as electronic chiralization (EC) due to resemblance to their optical counterparts \cite{lipkin_1964,tang_2010,bliokh_2011,coles_2012}, based on spatial gradients of periodic spin and charge densities of infinite crystals. In Sec.~\ref{sec:ECconstruction} we first introduce EC based on the symmetry properties of the AHE and then give convenient formulas for calculating EC in realistic systems. In Sec.~\ref{sec:ECexamples} we demonstrate EC's computation and behavior in several model examples: Anomalous Hall antiferromagnets Mn$_3$X (X = Ir, Pt, Sn, Ge etc.) and a 2D ferromagnetic Rashba model in plane wave basis. In Sec.~\ref{sec:new_AHE_systems} we give two nontrivial examples of the AHE inspired by the prominent role of magnetic charge in the EC: a minimal model based on the charge-ordered kagome spin ice \cite{wills_2002,moller_2009,chern_2011,qi_2008,ladak_2010,dun_2016,Matsuhira_2002,tabata_2006,takagi_1993,chern_2012,Wolf_1988,zhao_2020} and skew scattering of 2D Dirac electrons by magnetic charge. Finally in Sec.~\ref{sec:discussion} we briefly discuss further implications of the EC.

\section{Symmetry properties of the AHE and construction of the EC}\label{sec:ECconstruction}

\subsection{Symmetry properties of the AHE}\label{subsec:AHEsymmetry}

We first discuss the symmetry properties of the AHE of a crystal. The AHE is described by the anomalous Hall (pseudo)vector $(\boldsymbol{\sigma}_{\rm AH})_\alpha = \frac{1}{2}\epsilon_{\alpha\beta\gamma}\sigma_{\beta \gamma}$. $\boldsymbol{\sigma}_{\rm AH}$ changes sign under time reversal (TR, acting on the equilibrium state of the system) as a consequence of the Onsager relation; it rotates as a pseudovector under O(3) and is invariant under continuous translation operations because it is the response of a uniform current to a uniform electric field. 

It is obvious why $\boldsymbol{\sigma}_{\rm AH}$ must transform as a vector under proper rotations. By transform we mean the comparison between $\boldsymbol{\sigma}_{\rm AH}$ of two systems that are related to each other by such a transformation. Suppose that system 2 is obtained from system 1 by rigidly rotating the former. Then $\boldsymbol{\sigma}_{\rm AH}$ for system 2 is measured by applying the external electric field and attaching the current probes in the same manner as that for 1. In other words, we have the response relation 
\begin{eqnarray}
	\mathbf E \xrightarrow{\text{1}} \mathbf j_1,\;\mathbf E \xrightarrow{\text{2}} \mathbf j_2
\end{eqnarray} 
and the linear response function is defined as
\begin{eqnarray}
	\sigma_{1,2}^{\alpha\beta} = \frac{\partial j_{1,2}^\alpha}{\partial E_\beta}
\end{eqnarray}
However, because the rotation is equivalent to a coordinate transformation, we immediately know that if the electric field is rotated together with the system, the result should not change. In other words
\begin{eqnarray}
	R\mathbf E \xrightarrow{\text{2}} R\mathbf j_1 = \mathbf j_2.
\end{eqnarray}
where $R$ is a rotation matrix. Therefore
\begin{eqnarray}
	\sigma_{2}^{\alpha\beta} = \frac{\partial(R\mathbf j_1)^{\alpha}}{\partial(R\mathbf E)^\beta} = (R \sigma_1 R^{-1})^{\alpha\beta}.
\end{eqnarray}
It then follows that $\boldsymbol{\sigma}_{\rm AH, 2} = R \boldsymbol{\sigma}_{\rm AH, 1}$. A similar argument can be made for improper rotation and continuous translation. Alternatively one can use linear response theory and show that all unitary transformations due to O(3) operations cancel out when taking the trace.

In addition to the system-independent properties above, $\boldsymbol{\sigma}_{\rm AH}$ must also be invariant under any symmetry operations of the crystal, as dictated by Neumann's principle. In particular, for symmetry operations that combine a point group operation $R$ with a spatial translation $T$, since $\boldsymbol{\sigma}_{\rm AH}$ is invariant under continuous translation, it must be invariant under $R$ even if $R$ is not a symmetry of the crystal.

\subsection{Definition of the electronic chiralization}\label{subsec:ECdefinition}

Based on the discussion above, a suitable indicator of the AHE should be (1) a TR-odd pseudovector, and (2) invariant under all symmetry operations of the crystal. Then $\boldsymbol{\sigma}_{\rm AH}$ will be linearly dependent on this indicator to the lowest order of the latter. An important consequence of (1) is that the indicator must be translationally invariant. 

The TR-odd property of $\boldsymbol{\sigma}_{\rm AH}$ is fundamentally due to the microscopic magnetization density $\mathbf m( \mathbf r)$ in equilibrium. In ferromagnets the spatial average of $\mathbf m(\mathbf r)$, $\bar{\mathbf m}$ serves as a suitable indicator of the AHE. When $\bar{\mathbf m}$ nearly vanishes, it is reasonable to associate the AHE with the spatial variation of $\mathbf m(\mathbf r)$. We thus propose indicators of the AHE constructed from the spatial gradients of $\mathbf m(\mathbf r)$. For definiteness we only consider the first-order spatial derivative of $\mathbf m(\mathbf r)$ in this work, although indicators based on higher orders in $\mathbf m$ or its derivatives can be constructed similarly and may be useful in different cases. We start from a Cartesian tensor $T_{ijk}$ defined as
\begin{eqnarray}
T_{ijk} \equiv \frac{1}{V}\int d^3\mathbf r \partial_i \phi \partial_j m_k = \frac{1}{V_{\rm uc}}\int_{\rm uc} d^3\mathbf r \partial_i \phi \partial_j m_k
\end{eqnarray}
where $\phi$ is a TR-even scalar field observable of the crystal, which can be the charge density $\rho(\mathbf r)$ or the nonmagnetic potential $V(\mathbf r)$; uc stands for unit cell. We ignore any boundary contributions to $\mathbf m(\mathbf r)$ and $\phi(\mathbf r)$ so that they have the same discrete translation symmetry as the infinite crystal. The inclusion of $\phi$ ensures that $T_{ijk}$ does not become a boundary term and also signifies the role of orbital degrees of freedom in the AHE. $T_{ijk}$ is a TR-odd rank-3 pseudotensor and is translationally invariant. It is also invariant under any symmetry operations of the crystal since both $\phi(\mathbf r)$ and $\mathbf m(\mathbf r)$ are physical observables of the crystal. A pseudovector can be obtained from $T_{ijk}$ by contracting it with Kronecker $\delta$ or Levi-Civita symbol $\epsilon$. The only two independent pseudovectors obtained from this construction are
\begin{eqnarray}\label{eq:chi12}
&&\boldsymbol{\chi}_1 \equiv \frac{1}{V}\int d^3\mathbf r (\nabla \phi) (\nabla \cdot \mathbf m),\\\nonumber
&&\boldsymbol{\chi}_2 \equiv  \frac{1}{V}\int d^3\mathbf r (\nabla \phi) \times(\nabla \times \mathbf m).
\end{eqnarray}
We name $\boldsymbol{\chi}_{1,2}$ generally as ``electronic chiralization" to emphasize their electronic origin and pseudovector nature, analogous to the optical chirality (flow) in optics \cite{tang_2010,bliokh_2011,lipkin_1964,coles_2012}. Several comments are in order:

\noindent(i) One can define $\boldsymbol{\chi}_3 \equiv \frac{1}{V}\int d^3\mathbf r (\nabla^2 \phi) \mathbf m$ which is a linear combination of $\boldsymbol{\chi}_{1,2}$. However, a nonzero $\boldsymbol{\chi}_3$ in antiferromagnets suggests an AHE that is due to compensated $\mathbf m$ located on structurally inequivalent sites (different $\nabla^2\phi$) and is relatively trivial. We thus focus on $\boldsymbol{\chi}_1$ in this work only.  

\noindent(ii) $\boldsymbol{\chi}_1$ and $\boldsymbol{\chi}_2$ are respectively related to the magnetic charge density $\rho_m\equiv -\nabla \cdot \mathbf m$ and the electric current density $\mathbf j = \nabla \times \mathbf m$. When $\mathbf m$ can be approximated by $g\mu_{\rm B}\mathbf s(\mathbf r)$ with $\mathbf s(\mathbf r)$ the spin density, $\nabla \times \mathbf s(\mathbf r)$ is the ``spin current" contribution to the conserved charge current in the Dirac theory of electrons.

\noindent(iii) Using the electron charge density $\rho(\mathbf r)$ as the scalar field $\phi$, one can potentially obtain $\boldsymbol{\chi}_{1,2}$ directly from magnetic neutron or X-ray diffraction data since it only requires the knowledge of $\rho_{\mathbf K}^*\mathbf m_{\mathbf K} $, where $\mathbf K$ is a reciprocal lattice vector (see Sec.~\ref{subsec:ECformulas} below). Such a combination can appear, e.g. (for $\boldsymbol{\chi}_2$), in the interference term of elastic neutron scattering cross-section between magnetic and electrostatic scatterings \cite{lovesey_book}.

\noindent(iv) EC exists in ferromagnets as well, though in this case the net magnetization is a more straightforward indicator. In Sec.~\ref{subsec:EC_FMRashba} we show that in a modified Rashba model \cite{Bychkov_1984, Nagaosa_AHE_RMP_2010} the local spin and charge densities exhibit the characteristic textures that lead to nonzero $\boldsymbol{\chi}_{1,2,3}$.  

\noindent(v) One can generalize the above definition of EC by considering higher powers of $\phi$ or its derivative. For example:
\begin{eqnarray}\label{eq:EChigher}
	&&\boldsymbol{\chi}_1' =\frac{1}{V} \int d^3\mathbf r \phi (\nabla \phi) (\nabla \cdot \mathbf m), \\\nonumber
	&&\boldsymbol{\chi}_1'' =\frac{1}{V} \int d^3\mathbf r (\nabla \phi) (\nabla^2 \phi) (\nabla \cdot \mathbf m), \\\nonumber
	&&\dots
\end{eqnarray}
which are also nonzero in general if $\boldsymbol{\chi}_{1,2}\neq 0$. In a particular model it may be that $\boldsymbol{\chi}_{1,2}$ accidentally become zero due to certain artificial symmetry of the model (e.g. assuming a spherical charge/spin distribution for each atom). In such cases the alternative constructions above may be used. One example is hematite (Fe$_2$O$_3$) in the canted antiferromagnetic phase. We found that the EC calculated using the formula Eq.~\eqref{eq:chi12K} are zero, but $\boldsymbol{\chi}_1''$ is nonzero.

\subsection{Formulas for computing EC in realistic systems}\label{subsec:ECformulas} 

In this subsection we discuss how to efficiently calculate EC in realistic systems. Since crystallographic models are usually represented by localized atomic charge and magnetic moments, we consider such cases first by assuming that the atomic charge and spin densities are described by Gaussians, which gives
\begin{eqnarray}
	&&\rho(\mathbf r) = \sum_{\mathbf R}\sum_{n} Q_n g(\mathbf r-\mathbf R - \mathbf r_n)\\\nonumber
	&&\mathbf m(\mathbf r) = \sum_{\mathbf R}\sum_n \mathbf M_n g(\mathbf r-\mathbf R - \mathbf r_n)
\end{eqnarray}
where $Q_n$ and $\mathbf M_n$ are the charge and magnetic moment on a lattice site located at $\mathbf r_n$ in the unit cell, $\mathbf R$ stands for Bravais lattice vectors, and $g(\mathbf r) = (2\pi\sigma^2)^{-\frac{3}{2}} e^{-\frac{r^2}{2\sigma^2}}$ is the Gaussian function whose Fourier transform is $g_{\mathbf k} = e^{-\frac{\sigma^2 k^2}{2}}$. The Fourier transform of $\rho(\mathbf r)$ and $\mathbf m(\mathbf r)$ are therefore
\begin{eqnarray}
	&&\rho_{\mathbf K} = \frac{1}{V_{\rm uc}} \sum_n Q_n g_{\mathbf K} e^{-\imath \mathbf K \cdot \mathbf r_n},\\\nonumber
	&&\mathbf m_{\mathbf K} = \frac{1}{V_{\rm uc}} \sum_n \mathbf M_n g_{\mathbf K} e^{-\imath \mathbf K \cdot \mathbf r_n},
\end{eqnarray}
from which we can obtain $\boldsymbol{\chi}_{1,2}$
\begin{eqnarray}\label{eq:chi12K}
	&&\boldsymbol{\chi}_1 = \sum_{\mathbf K} \mathbf K \rho^*_{\mathbf K} (\mathbf K \cdot \mathbf m_{\mathbf K}) = \frac{1}{V_{\rm uc}^2} \sum_{\mathbf K} g_{\mathbf K}^2 \mathbf K (\mathbf K\cdot \mathbf X_{\mathbf K})\\\nonumber
	&&\boldsymbol{\chi}_2 = \sum_{\mathbf K} \mathbf K \rho^*_{\mathbf K} \times (\mathbf K \times \mathbf m_{\mathbf K}) = \frac{1}{V_{\rm uc}^2} \sum_{\mathbf K} g_{\mathbf K}^2 \mathbf K \times (\mathbf K\times \mathbf X_{\mathbf K})
\end{eqnarray}
where 
\begin{eqnarray}
	\mathbf X_{\mathbf K} \equiv \sum_{mn}Q_m \mathbf M_{n} e^{\imath \mathbf K \cdot (\mathbf r_m - \mathbf r_{n})}.
\end{eqnarray}
Because of the $g_{\mathbf K}^2$ in Eq.~\eqref{eq:chi12K} the summation will converge quickly if $\sigma$ is comparable to the lattice constant.

Although the momentum space expressions above can be directly applied to scattering data (with the $g_{\mathbf K}$ replaced by appropriate form factors) and plane-wave-based DFT calculations (see Sec.~\ref{subsec:EC_FMRashba} for an example), they do not necessarily provide a transparent picture on the real-space charge and magnetization distributions. Moreover, when the charge and magnetization densities are very localized around individual atoms, and have fine structures at small length scales (or equivalently at high energy scales) that can be irrelevant to transport phenomena, the values of $\boldsymbol{\chi}_{1,2}$ can be sensitive to the cutoff in $\mathbf K$. We therefore discuss real-space expressions of $\boldsymbol{\chi}_{1,2}$ next, still assuming that the local densities for each atom are described by Gaussians.

For $\boldsymbol{\chi}_1$ we have
\begin{flalign}\label{eq:chi1realspace}
	&\boldsymbol{\chi}_1 = &\\\nonumber
	&\frac{1}{V_{\rm uc}}\sum_{mn}\sum_{\mathbf R}\int d^3\mathbf r Q_n \mathbf M_m\cdot [\nabla g(\mathbf r -  \mathbf r_m)] [\nabla g(\mathbf r - \mathbf R - \mathbf r_n)]&\\\nonumber
	&\equiv \frac{1}{V_{\rm uc}}\sum_{mn}\sum_{\mathbf R} Q_n \mathbf M_m\cdot \overleftrightarrow{I}(\mathbf R + \mathbf r_n - \mathbf r_m)&
\end{flalign}
in which the integral
\begin{eqnarray}
	I_{ij}(\mathbf a) &=& -\partial_{a_i}\partial_{a_j} \int d^3 \mathbf r g(\mathbf r) g(\mathbf r + \mathbf a)\\\nonumber
&=&\frac{1}{16\pi^{\frac{3}{2}}\sigma^5}\left( \delta_{ij} - \frac{a_i a_j}{2\sigma^2} \right)e^{-\frac{a^2}{4\sigma^2}}.
\end{eqnarray}
Because of the Gaussian factor in $\overleftrightarrow{I}$ one can potentially truncate the summation in Eq.~\eqref{eq:chi1realspace} at e.g. nearest neighbor. Note that the second term in $\overleftrightarrow{I}$ has the form of an electric quadrupole. The main difference between it and the quadrupole moment in \cite{Hayami_AHE_AMD_2021} is that the origin of the former is different for different pairs of ions.

Using the same $\overleftrightarrow{I}$ one can express $\boldsymbol{\chi}_{2}$ as
\begin{eqnarray}
	\chi_{2k} &=& \frac{1}{V_{\rm uc}}\sum_{mn}\sum_{\mathbf R} Q_n M^b_m \epsilon_{ijk}\epsilon_{abj} I_{ai}(\mathbf R + \mathbf r_{mn}) \\\nonumber
	&=& \frac{1}{V_{\rm uc}}\sum_{mn}\sum_{\mathbf R} Q_n M^b_m \left( I_{bk} - \delta_{bk} {\rm Tr}\overleftrightarrow{I}\right).
\end{eqnarray}
The contributions due to the traceless part of $\overleftrightarrow{I}$ to $\boldsymbol{\chi}_{1,2}$ are therefore the same. The trace of $\overleftrightarrow{I}$ gives rise to a weighted sum of electric charge surrounding a given magnetic moment $\mathbf M_m$. A compensated ferrimagnet will have a nonzero $\boldsymbol{\chi}_1$ mainly due to this contribution. We will focus the traceless part of $\overleftrightarrow{I}$ in Sec.~\ref{sec:ECexamples} below since it is less trivial. This also makes it sufficient to consider $\boldsymbol{\chi}_1$ only. Real-space formulas for the generalized EC introduced in Eq.~\eqref{eq:EChigher} are given in Appendix~\ref{appendix:gEC}.

\section{EC in model examples}\label{sec:ECexamples}

In this section we calculate the EC using the formulas given in the last section and discuss their connections with the AHE in several model examples.

\subsection{EC in AHE antiferromagnets}\label{subsec:EC_AHEAFM}

We first calculate $\bm \chi_1$ using crystallographic models for the noncollinear AHE antiferromagnets Mn$_3$X (X = Ir, Pt, Sn, Ge, etc.) and the real-space formula Eq.~\eqref{eq:chi1realspace}. Considering nearest neighbors and the traceless part of $\overleftrightarrow{I}$ only, we have $\boldsymbol{\chi}_1 = C\sum_{m} \mathbf M_{m} \cdot \mathcal{Q}_m$, where $C = - e^{\frac{r_{\rm nn}^2}{4\sigma^2}}/(32 \pi^{\frac{3}{2}}\sigma^7 V_{\rm uc})$ is a system dependent constant and $r_{\rm nn}$ is the distance between a magnetic atom and its nearest neighbors. $\mathcal{Q}_m$ is the total electric quadrupole relative to the position of site $m$:
\begin{eqnarray}
	\mathcal{Q}_m = \sum_{i\in \{i\}_m} Q_i \left(\mathbf r_{mi} \mathbf r_{mi}-\frac{1}{3}r_{\rm nn}^2  \mathbb{I} \right)
\end{eqnarray}
where $\{i\}_m$ stands for the set of nearest neighbors of site $m$. $\mathcal{Q}_m$ has the same symmetry as a second-order, i.e. easy-axis or easy-plane, magnetic anisotropy. The origin of weak ferromagnetism and hence the AHE in systems with a nonzero $\boldsymbol{\chi}_1$ is thus a site-dependent second order anisotropy, which applies to the known examples of Mn$_3$X and collinear AHE antiferromagnets \cite{Smejkal_CHE_2020, Li_QAHE_2019}. If such a site-dependent second order anisotropy is forbidden by symmetry, as in the case of hematite, one needs to consider the generalizations of EC as mentioned in Sec.~\ref{subsec:ECformulas}. 

\begin{figure}[ht]
	\centering
	\subfloat[]{\includegraphics[width=1.1 in]{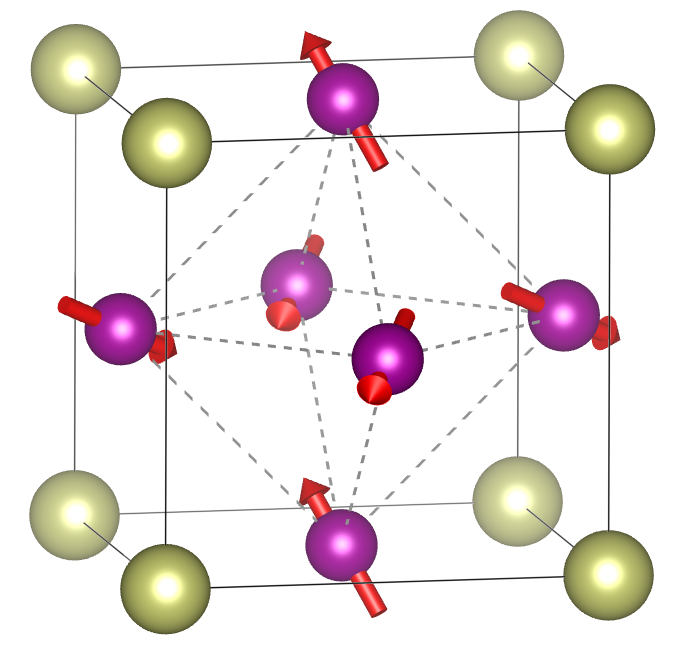}}\;\;
	\subfloat[]{\includegraphics[width=1.6 in]{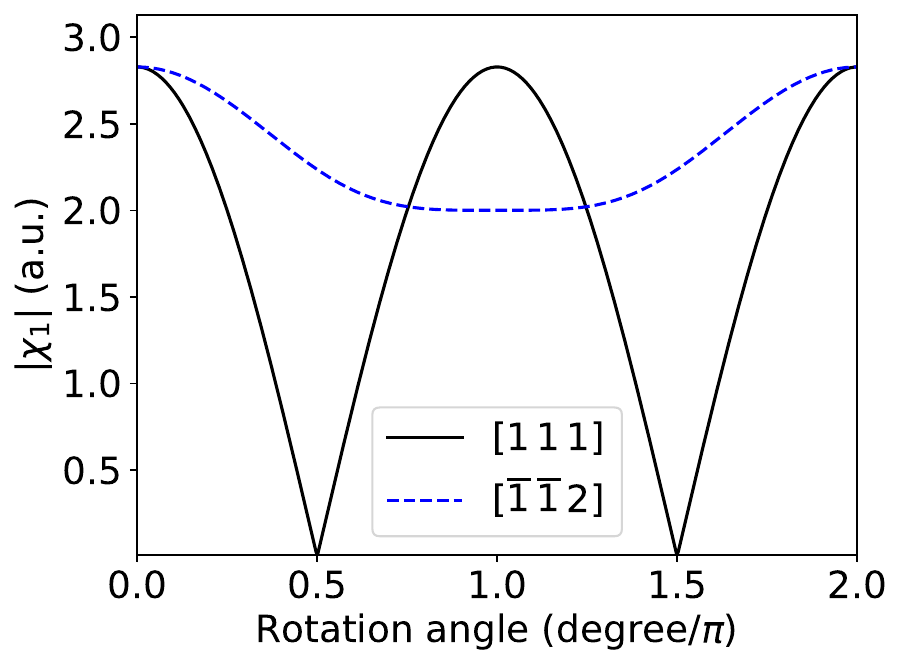}}\\
	\subfloat[]{\includegraphics[width=1.45 in]{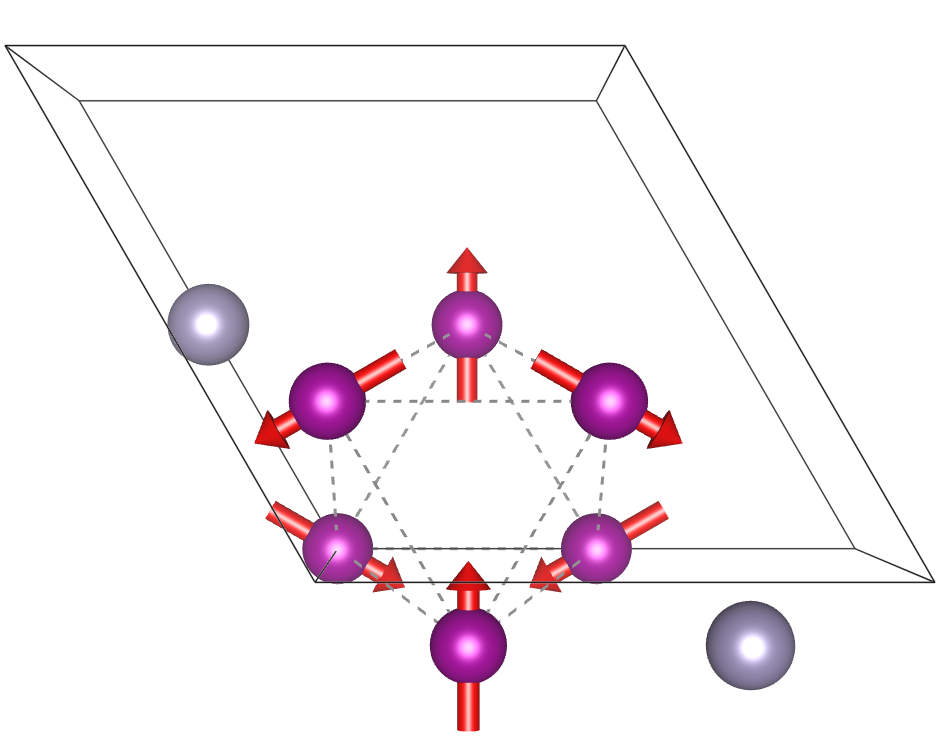}}
	\subfloat[]{\includegraphics[width=1.6 in]{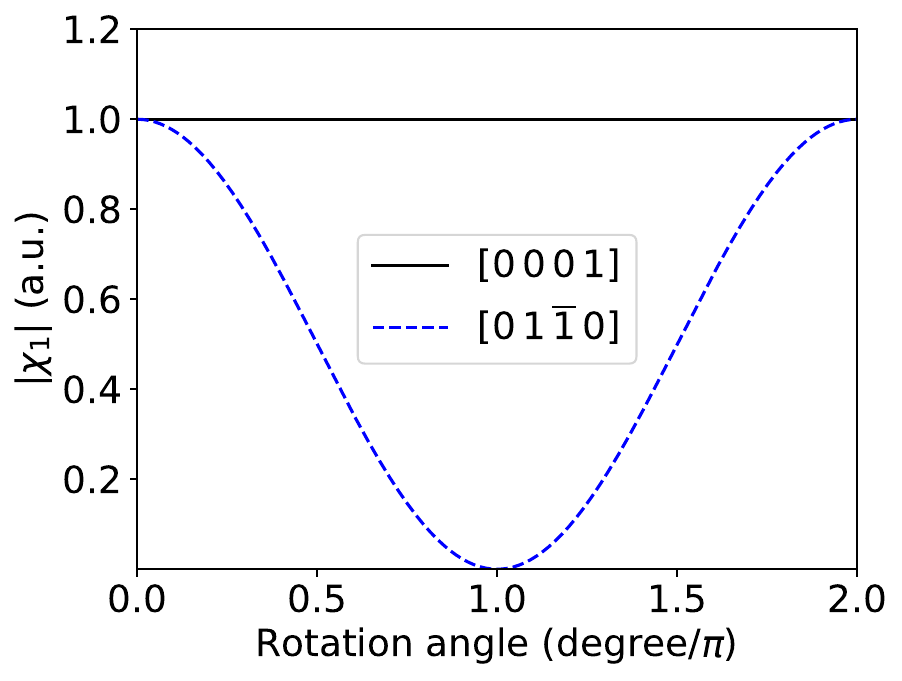}}
	\caption{(Color online) (a) Crystal structure and magnetic order of cubic Mn$_3$X. (b) Dependence of $|\boldsymbol{\chi}_1|$ on rigid rotation of all sublattice magnetic moments about $[111]$ (perpendicular to the initial kagome plane that all moments are parallel with) and $[\bar{1}\bar{1}2]$ (along the initial direction of a sublattice moment), respectively. (c) Crystal structure and magnetic order of hexagonal Mn$_3$X. (d) Dependence of $|\boldsymbol{\chi}_1|$ on rigid rotation of all sublattice magnetic moments about $[0001]$ (perpendicular to the kagome plane) and $[01\bar{1}0]$ [along the initial direction of the moment on the topmost atom in (c)], respectively.} 
	\label{fig:ecmodel}
\end{figure}

For the structure of cubic Mn$_3$X, e.g. Mn$_3$Ir [Fig.~\ref{fig:ecmodel} (a)], $\mathcal{Q}_m$ is diagonal with the principal axis of the largest eigenvalue along the four-fold axis on each Mn atom. One can then obtain the dependence of $\boldsymbol{\chi}_1$ on rigid rotations of the sublattice moments, which is very different from rotating the pseudovector $\boldsymbol{\chi}_1$ directly since the latter is not required to transform as a pseudovector under separate rotations of the lattice and magnetic moments. For example, the length of $\boldsymbol{\chi}_1$ depends on the rotation about its direction as $|\cos\gamma|$ [Fig.~\ref{fig:ecmodel} (b)], where $\gamma$ is the rotation angle, similar to $\boldsymbol{\sigma}_{\rm AH}$ (see below) and the orbital magnetization \cite{Chen_2020}. For the structure of hexagonal Mn$_3$X, e.g. Mn$_3$Sn [Fig.~\ref{fig:ecmodel} (c)], one can follow the same procedure and obtain a compact expression: 
\begin{eqnarray}
\boldsymbol{\chi}_1 \propto \cos^2\left( \frac{\beta}{2}  \right)\left[ \sin(\alpha+\gamma) \hat{x} +  \cos(\alpha+\gamma) \hat{y} \right]
\end{eqnarray}
where $\alpha,\beta,\gamma$ are Euler angles about the $z,y',z''$ axes, respectively. Here $z$ and $y$ are the original crystalline axes $[0001]$ and $[01\bar{1}0]$, while the primed and double-primed axes correspond to those co-rotated with the local moments successively by $\alpha$ and $\beta$, respectively. The expression captures the phenomenon that a counterclockwise rotation about $[0001]$ of all sublattice moments leads to clockwise rotation of the weak magnetization and $\boldsymbol{\sigma}_{\rm AH}$ \cite{Nakatsuji_2015}. Moreover, rotation about the $[01\bar{1}0]$ axis in Fig.~\ref{fig:ecmodel} (c) by $\pi$ makes $\boldsymbol{\chi}_1$ vanish [Fig.~\ref{fig:ecmodel} (d)], since in this case the magnetic order becomes triangular rather than inverse triangular, and the AHE or weak ferromagnetism are forbidden by a $C_3$ symmetry.

To compare the angular dependence of $|\bm \chi_1|$ with that of $|\boldsymbol{\sigma}_{\rm AH}|$, we use the following generic tight-binding model \cite{Chen_2020} adapted to the structures of cubic and hexagonal Mn$_3$X to calculate the intrinsic contribution to the anomalous Hall conductivity:
\begin{flalign}\label{eq:TBmodel}
	&H = H_t + H_{\rm so} + H_J &\\\nonumber
	&\equiv -t\sum_{\langle i j\rangle \alpha} c_{i\alpha}^\dag c_{j\alpha} + \imath \lambda_{\rm so} \sum_{\langle ij\rangle \alpha\beta} (\hat{r}_{ij}\times \boldsymbol{\eta}_{ij})\cdot \boldsymbol{\sigma}_{\alpha\beta} c_{i\alpha}^\dag c_{j\beta} &\\\nonumber
	&- J\sum_{i\alpha\beta} \hat{n}_i \cdot \boldsymbol{\sigma}_{\alpha\beta} c_{i\alpha}^\dag c_{i\beta}&
\end{flalign}
where $i,j$ label lattice sites, $\langle \rangle $ means nearest neighbor, $\alpha,\beta$ label spin, $t >0$ is the spin-independent hopping amplitude and is chosen as the energy unit, $\lambda_{\rm so}$ is the spin-orbit coupling strength, $\hat{r}_{ij}$ is a unit vector along the position vector $\mathbf r_j - \mathbf r_i$, $\boldsymbol{\eta}_{ij}$ is the electric field or electric dipole moment vector at the center of the nearest-neighbor $ij$ bond \cite{Chen_2020} (normalized using the largest $|{\eta}_{ij}|$), $J$ is the strength of a local exchange field along $\hat{n}_i$ mimicking the noncollinear magnetic order in a given material.

The anomalous Hall conductivity is calculated as 
\begin{flalign}
	&\sigma_{\rm AH}^{\gamma} = \frac{1}{2}\epsilon^{\alpha\beta\gamma}\sigma^{\alpha\beta}(\omega = 0)&\\\nonumber
	& = \frac{e^2\hbar \epsilon^{\alpha\beta\gamma}}{2N_{\mathbf k} V_{\rm uc}} \sum_{n\neq m; \mathbf k} \frac{f_{n \mathbf k} - f_{m \mathbf k}}{ (\epsilon_{n \mathbf k} - \epsilon_{m \mathbf k})^2 + \eta^2 } {\rm Im}\left( v^\alpha_{nm\mathbf k} v^{\beta}_{mn\mathbf k}  \right)&
\end{flalign}
where $\sigma^{\alpha\beta}(\omega)$ is the optical conductivity tensor, $N_{\mathbf k}$ is the number of points of the $k$-mesh, and $\eta$ is a band broadening parameter that depends on disorder. To facilitate rapid convergence of the Brillouin zone integration we have used $\eta = 0.1$ (in units of $t$) and a thermal smearing with $k_{\rm B} T = 0.3$. Such smearing parameters also help to eliminate any spurious abrupt changes of the zero-temperature intrinsic AHC in a perfect crystal versus smooth changes of tuning parameters such as the rotation angle, since the former is sensitively dependent on small band splittings at the Fermi energy. In reality the dependence of transport coefficients on orientations of the magnetic order parameter are expected to consist of low-order Fourier components due to both thermal and disorder effects. The Brillouin zone integration is performed using a $31\times 31\times 31$ mesh.

\begin{figure}[ht]
	\centering
	\subfloat[]{\includegraphics[width=1.65 in]{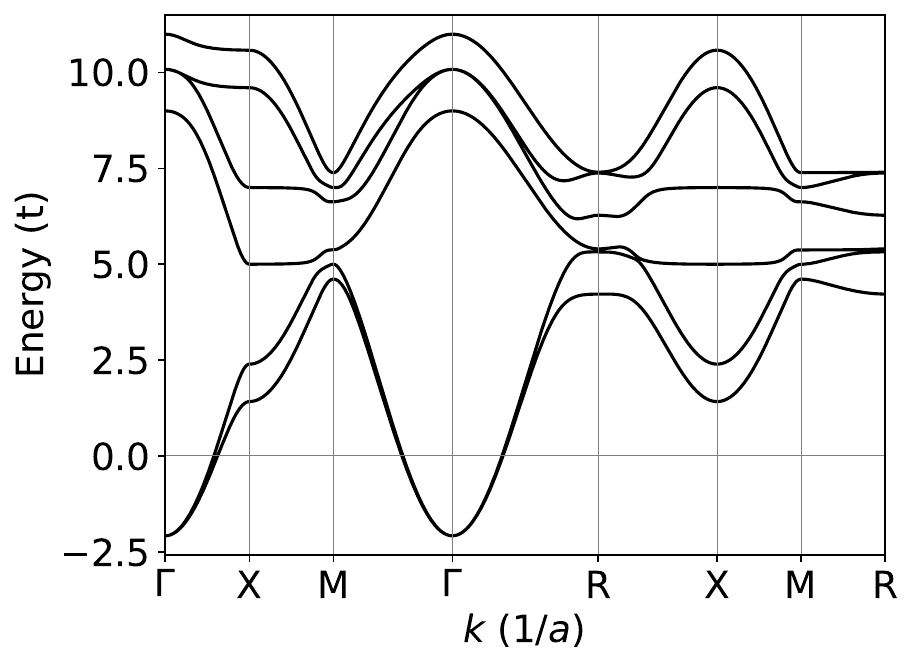}}
	\subfloat[]{\includegraphics[width=1.55 in]{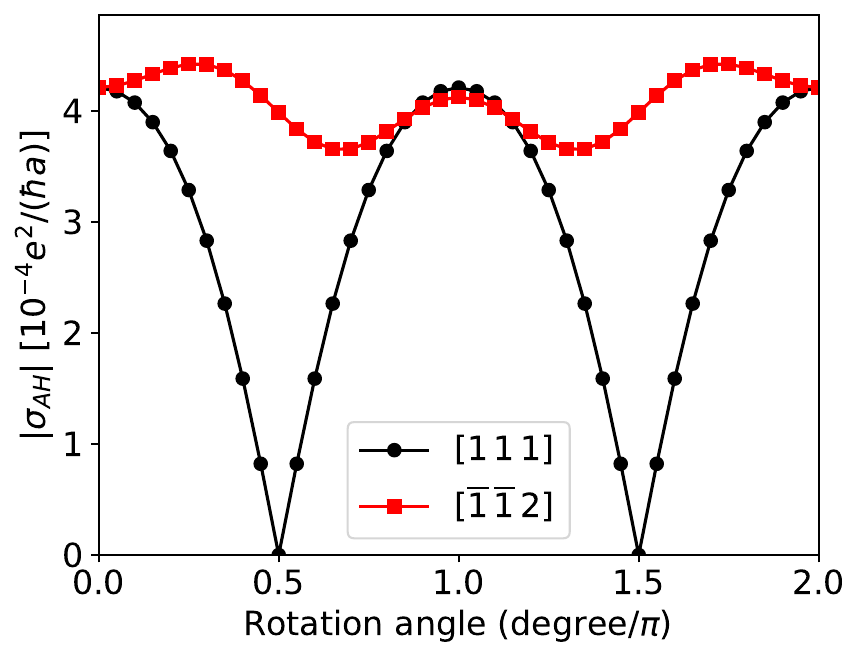}}\\
	\subfloat[]{\includegraphics[width=1.6 in]{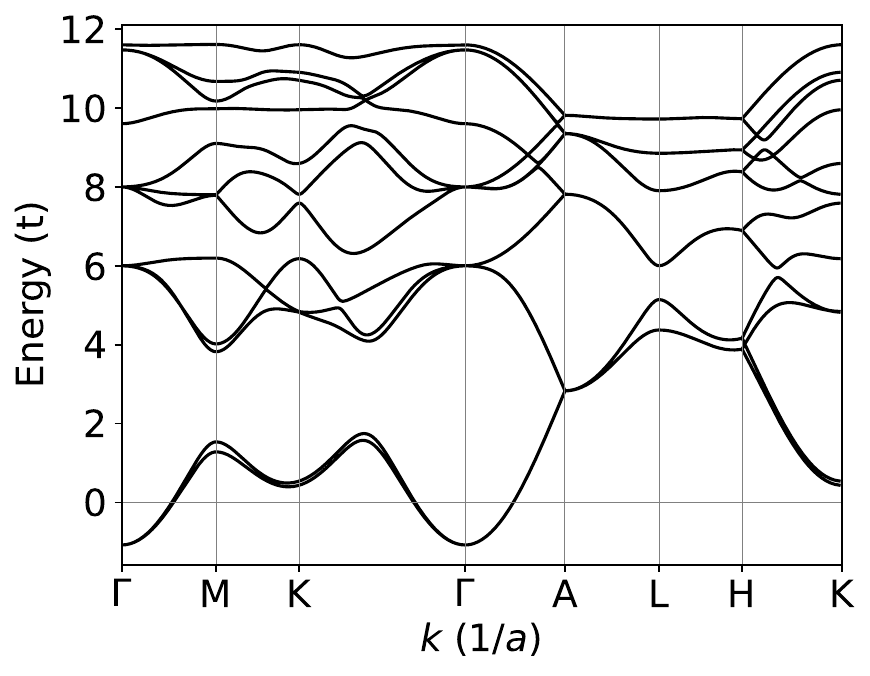}}
	\subfloat[]{\includegraphics[width=1.65 in]{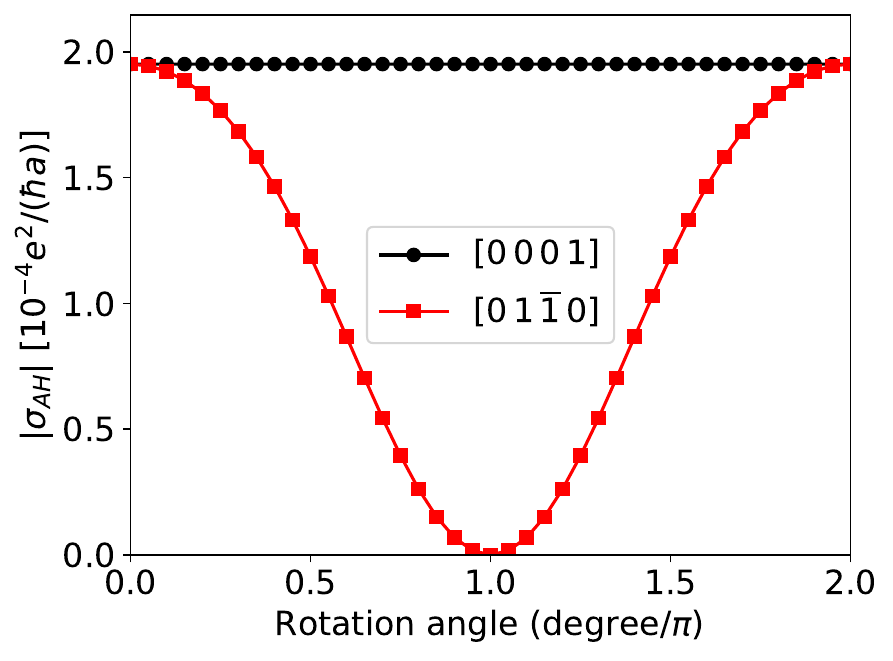}}
	\caption{(Color online) (a) Band structure of the model in Eq.~\eqref{eq:TBmodel} for the cubic Mn$_3$X structure, obtained using $t=1,\, \lambda_{\rm so} = 0.1, \, J = 1,\, \mu = -6$, where $\mu$ is the chemical potential. (b) Angular dependence of the norm of the AHC vector for the cubic Mn$_3$X model. (c) Band structure for the hexagonal Mn$_3$X structure, obtained using $t=1,\, \lambda_{\rm so} = 0.2, \, J = 1,\, \mu = -7$. (d) Angular dependence of the norm of the AHC vector for the hexagonal Mn$_3$X model.} 
	\label{fig:ahcvsrot}
\end{figure}

Figures.~\ref{fig:ahcvsrot} (b) and (d) agree qualitatively with the angular dependence of the EC in Figs.~\ref{fig:ecmodel} (b) and (d). Such an agreement to some extent depends on the parameter values used, but is generally expected based on symmetry arguments: For an arbitrary rotation of the local magnetic moments along a closed path one can generally expand the $\boldsymbol{\sigma}_{\rm AH}$ and EC as Fourier series. The high-symmetry points on the rotation path, either corresponding to exact vanishing of the $\boldsymbol{\sigma}_{\rm AH}$ and EC (e.g., $\pi$ rotation about $[111]$ in Mn$_3$Ir), or equivalent to certain magnetic space group operations (e.g., $2\pi/3$ rotation about $[0001]$ in Mn$_3$Sn), place identical constraints on the Fourier coefficients of $\boldsymbol{\sigma}_{\rm AH}$ and EC. As a result, when the angular dependence is smooth so that only a few low-order Fourier coefficients are relevant, it is expected that $\boldsymbol{\sigma}_{\rm AH}$ and EC should behave similarly. Such a symmetry analysis is analogous to that commonly used in ferromagnetic crystals \cite{birss1964symmetry}. Nonetheless, $\boldsymbol{\sigma}_{\rm AH}$ and EC are not required to have the same angular dependence, which is also similar to the relation between $\boldsymbol{\sigma}_{\rm AH}$ and the net magnetization in a ferromagnet. 

\subsection{EC in a ferromagnetic Rashba model}\label{subsec:EC_FMRashba}

\begin{figure}[ht]
	\centering
	\subfloat[]{\includegraphics[width=1.55 in]{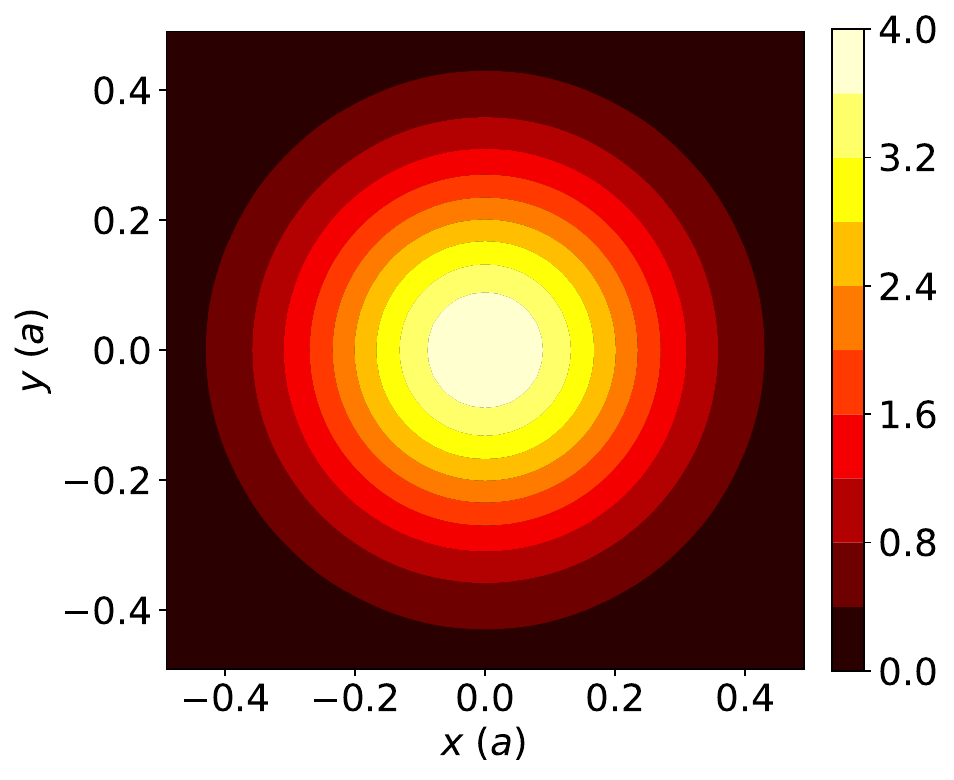}}
	\subfloat[]{\includegraphics[width=1.65 in]{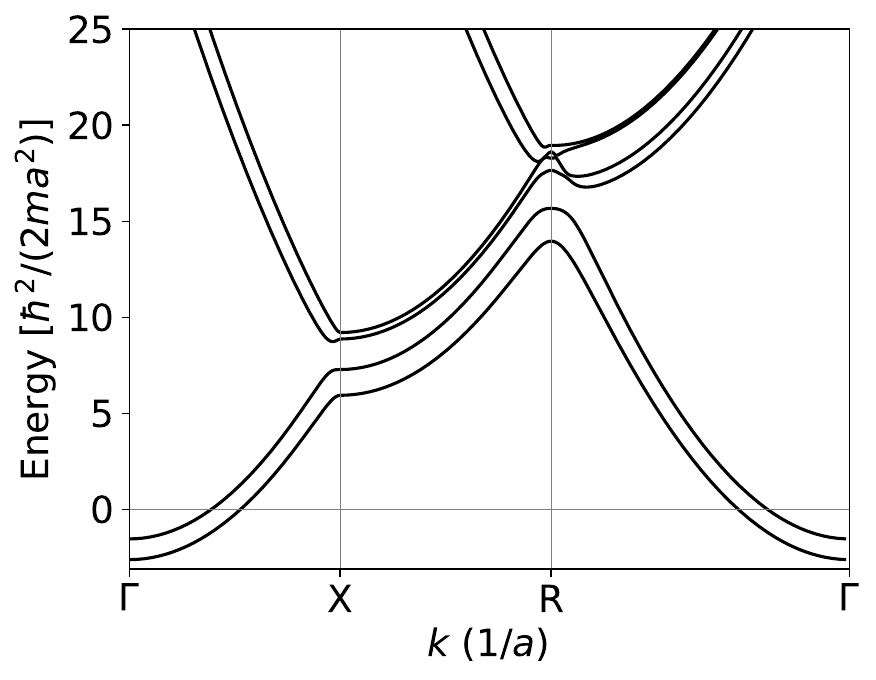}}\\
	\subfloat[]{\includegraphics[width=1.7 in]{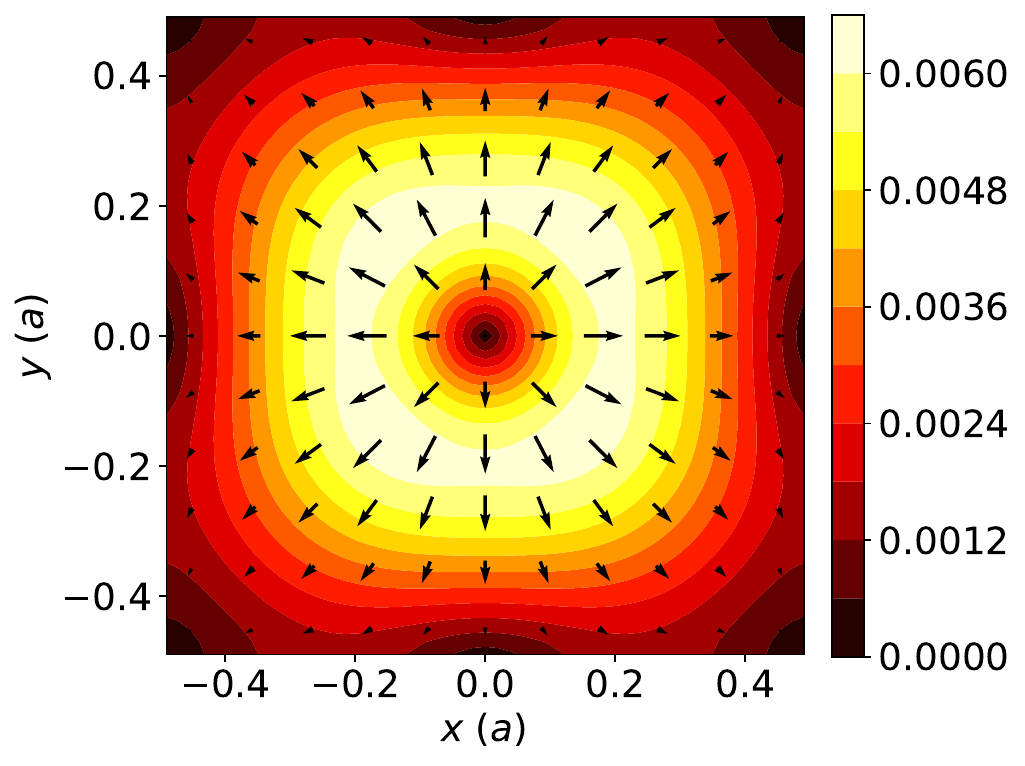}}
	\subfloat[]{\includegraphics[width=1.65 in]{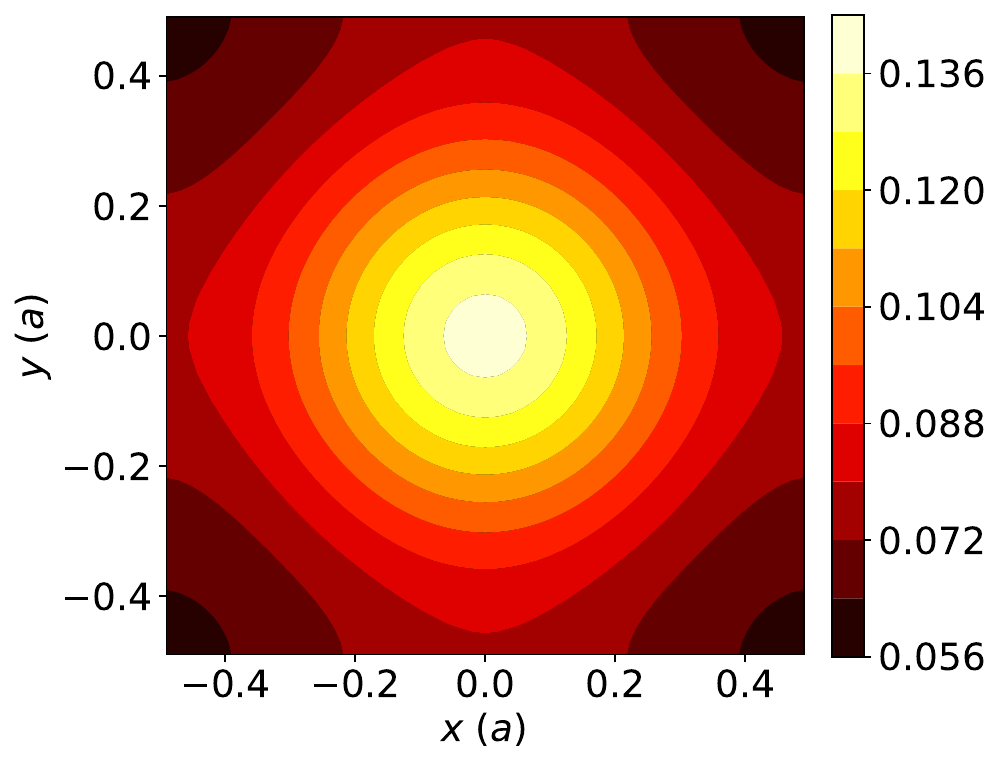}}\\
	\subfloat[]{\includegraphics[width=1.6 in]{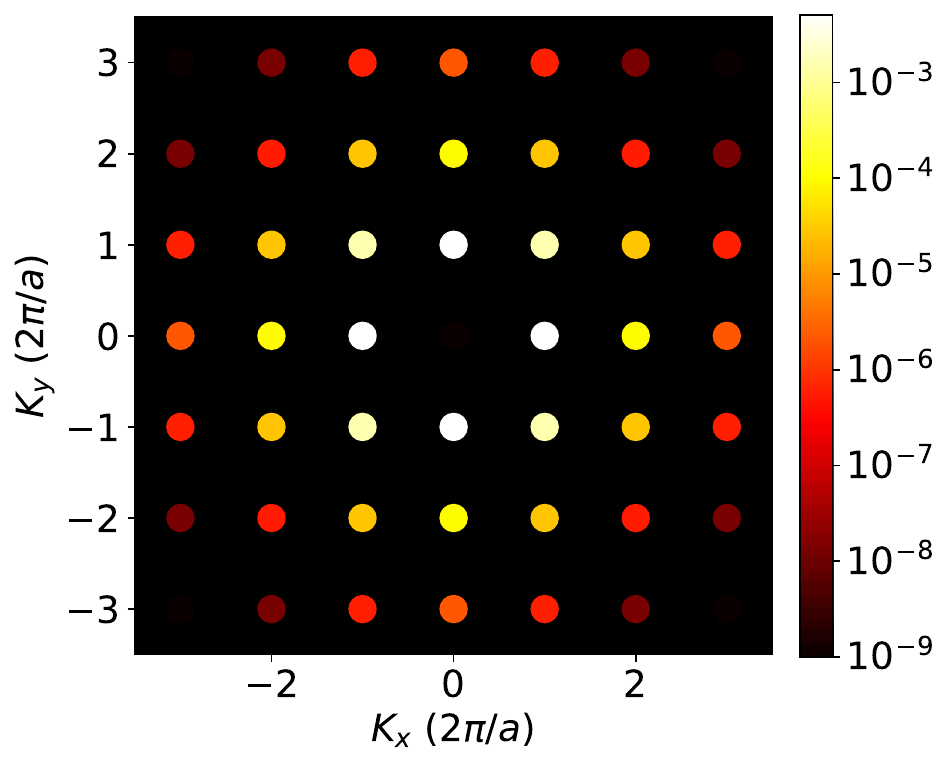}}
	\subfloat[]{\includegraphics[width=1.65 in]{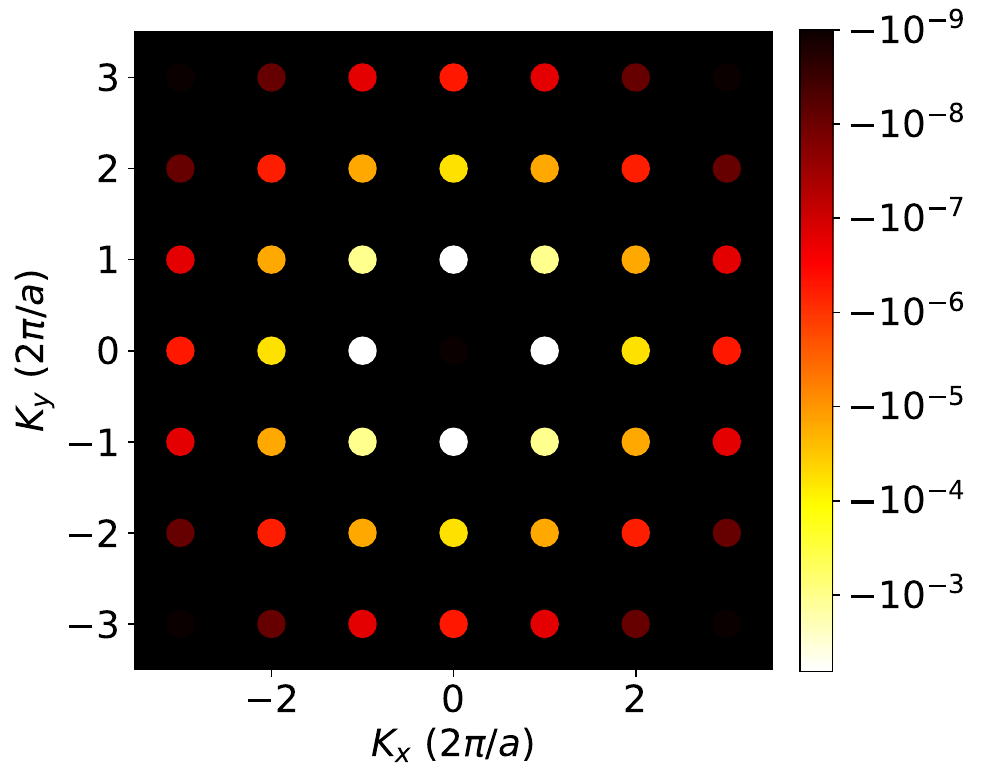}}	
	\caption{(Color online) (a) Spatial profile of the periodic confinement potential $G(\mathbf r)$ in Eq.~\eqref{eq:Grconfine}; $K_G = 10\left( \frac{2\pi}{a}\right)$. (b) Band structure along the high-symmetry lines in the Brillouin zone. The parameter values are: $t = 1,\, \lambda_{R} = 0.2,\, J = 0.5, \, V= 2,\, \mu = 0,\, K_H = K_G = 10\left( \frac{2\pi}{a}\right)$. Only the 6 lowest bands are shown. (c) In-plane components of the spin density in the unit cell. The color coding represents the size of the local in-plane spin density in units of $\frac{\hbar}{2 a^2}$. The arrows represent both the size and direction of the spin density. (d) Out-of-plane component of the spin density in the unit cell. (e) Bragg peaks corresponding to the summand of $\chi_1^z$ in Eq.~\eqref{eq:chi1z} plotted in log scale. (f) Bragg peaks for $\chi_2^z$ in Eq.~\eqref{eq:chi2z}.} 
	\label{fig:FMmodel}
\end{figure}

In this subsection we give an example of a microscopic calculation of $\boldsymbol{\chi}_1$ by using a 2D Rashba-type continuum model. The procedure can be straightforwardly applied to plane-wave-based density functional theory calculations. The model also serves as an example of the applicability of EC for conventional ferromagnets. The Hamiltonian describes itinerant electrons subject to periodic scalar, Zeeman, and Rashba spin-orbit coupling potentials on a square lattice:
\begin{flalign}\label{eq:HRashbaconf}
	&H = -\frac{\hbar^2}{2 m}\nabla^2 + \frac{\imath}{2}\lambda_{R}  \{G(\mathbf r),\hat{z}\cdot(\boldsymbol{\sigma} \times \nabla)\} - J G(\mathbf r) \sigma_z&\\\nonumber
	& - V G(\mathbf r),&
\end{flalign}
where $\lambda_R$ is the strength of the Rashba spin-orbit coupling, $J$ is that of a Zeeman field along $\hat{z}$, and $V$ is that of a scalar confinement potential. $\{,\}$ stands for anti-commutation and is needed to ensure that the position-dependent Rashba term is Hermitian. $G(\mathbf r)$ is a periodic Gaussian-like function:
\begin{eqnarray}\label{eq:Grconfine}
	G(\mathbf r) = \frac{1}{V_{\rm uc}}\sum_{\mathbf K} \int_{\rm uc} d^2\mathbf r' g(\mathbf r') e^{\imath \mathbf K\cdot(\mathbf r - \mathbf r')} \equiv \sum_{\mathbf K} G_{\mathbf K} e^{\imath \mathbf K\cdot \mathbf r},
\end{eqnarray}
where uc stands for unit cell defined by the lattice vectors $\mathbf a_1 = a \hat{x},\mathbf a_2 = a\hat{y}$, $\mathbf K$ are reciprocal lattice vectors, and $g(\mathbf r)$ is the 2D Gaussian $g(\mathbf r) = \frac{1}{2\pi \sigma^2} e^{-\frac{r^2}{2\sigma^2}}$. In practice we will set $G_{\mathbf K} = 0$ when $\max(|K_x|,|K_y|)>K_G$, $K_G$ being a parameter. We choose $a$ as the length unit, and $E_0 = \hbar^2 /(2ma^2) $ as the energy unit. An illustration of $G(\mathbf r)$ with $K_G = 10\left( \frac{2\pi}{a}\right)$ is shown in Fig.~\ref{fig:FMmodel} (a). In the plane wave basis an arbitrary eigenfunction can be written as
\begin{eqnarray}
	\psi = \sum_{\mathbf k} c_{\mathbf k} e^{\imath \mathbf k \cdot \mathbf r} \equiv \sum_{\mathbf k\in {\rm BZ}} \sum_{\mathbf K} c_{\mathbf K} (\mathbf k) e^{\imath (\mathbf k+\mathbf K) \cdot \mathbf r},
\end{eqnarray}
where $c$ are $2\times 1$ column vectors. Substituting this wavefunction into the eigen-equation $H\psi = \epsilon \psi$ for the dimensionless Hamiltonian and using the orthogonality between plane waves we obtain 
\begin{flalign}\label{eq:HRashbamat}
	&(\mathbf k+ \mathbf K)^2 c_{\mathbf K}(\mathbf k) + \sum_{\mathbf K'} G_{\mathbf K - \mathbf K'} U_{\mathbf K \mathbf K'}(\mathbf k) c_{\mathbf K'} (\mathbf k)  &\\\nonumber
	&  - \epsilon c_{\mathbf K} (\mathbf k) = 0&
\end{flalign}
where 
\begin{flalign}
&	U_{\mathbf K \mathbf K'} (\mathbf k) \equiv - J\sigma_z - V &\\\nonumber
&+ \lambda_R \left[\left(k_x + \frac{K_x + K'_x}{2}\right)\sigma_y - \left(k_y + \frac{K_y + K'_y}{2}\right) \sigma_x\right].&
\end{flalign}

Equation~\eqref{eq:HRashbamat} represents infinite coupled linear equations for a given $\mathbf k$, or a matrix equation for the column vector $c_{\mathbf K}(\mathbf k)$, with $\mathbf K$ understood as a row or column index. We truncate the Hamiltonian matrix by requiring $\max(|K_x|, |K_y|) \leq K_H$. Note that $K_H$ is generally different from $K_G$ defined above. The dimension of the Hamiltonian matrix for our model is therefore $N\times N$, $N = 2(2K_H + 1)^2$ ($K_H$ is in units of $2\pi/a$). Figure~\ref{fig:FMmodel} (b) shows a typical band structure of the model [$K_H = 10\left( \frac{2\pi}{a}\right)$, only the lowest six bands are plotted]. 

The spin density (at position $\boldsymbol{\tau}$) operator has the following matrix elements (taking $\frac{\hbar}{2}$ as the units of spin)
\begin{eqnarray}\label{eq:smatKKptau}
	\mathbf s_{\mathbf K \mathbf K'}(\boldsymbol{\tau}) = \boldsymbol{\sigma} e^{-\imath(\mathbf K - \mathbf K')\cdot \boldsymbol{\tau}}
\end{eqnarray}
whose Fourier transform at reciprocal lattice vector $\mathbf K_0$ is
\begin{eqnarray}\label{eq:smatKKpK0}
	\mathbf s_{\mathbf K \mathbf K'}(\mathbf K_0) = \boldsymbol{\sigma} \delta_{\mathbf K_0, \mathbf K' - \mathbf K}
\end{eqnarray}
The expectation value of the spin density for a given chemical potential is therefore
\begin{eqnarray}\label{eq:sdentau}
	\mathbf s (\boldsymbol{\tau}) = \sum_{n \mathbf k}\langle n \mathbf k | \mathbf s(\boldsymbol{\tau}) | n \mathbf k \rangle f_{n \mathbf k}
\end{eqnarray}
where $f_{n \mathbf k}$ is the Fermi-Dirac distribution function at eigen-energy $\epsilon_{n\mathbf k}$. Its Fourier transform at reciprocal lattice vector $\mathbf K_0$ is 
\begin{eqnarray}\label{eq:sdenK0}
	\mathbf s_{\mathbf K_0} = \sum_{n \mathbf k} \langle n \mathbf k | \mathbf s(\mathbf K_0) | n \mathbf k \rangle f_{n \mathbf k}.
\end{eqnarray}

The ferromagnetic Rashba model \cite{Bychkov_1984} is known to have the AHE with an out-of-plane $\boldsymbol{\sigma}_{\rm AH}$ \cite{Nagaosa_AHE_RMP_2010}. Therefore we only consider the out-of-plane components of the EC. Taking the spatial dependence of the Rashba coefficient as $\partial_z \phi(\mathbf r)$ in the definition of $\boldsymbol{\chi}_1$, we can finally obtain 
\begin{eqnarray}\label{eq:chi1z}
	\chi_1^z = \imath \sum_{\mathbf K} G^*_{\mathbf K} \mathbf K\cdot \mathbf s_{\mathbf K}.
\end{eqnarray}
Moreover, the spatial dependence of the Gaussian potentials allows us to calculate $\boldsymbol{\chi}_2$ and $\boldsymbol{\chi}_3$ as well, for which we will directly use the electron density $\rho(\boldsymbol{\tau})$ as the scalar field, whose expressions are similar to Eqs.~\eqref{eq:smatKKptau}--\eqref{eq:sdenK0} but with $\boldsymbol{\sigma}$ replaced by $\sigma_0$. Then
\begin{eqnarray}\label{eq:chi2z}
	\chi_2^z = -\sum_{\mathbf K} K^2 \rho_{\mathbf K}^* s^z_{\mathbf K} = \chi_3^z.
\end{eqnarray}
Note that in the present case $\boldsymbol{\chi}_3$ is not a linear combination of $\boldsymbol{\chi}_{1,2}$ any more.

Figures~\ref{fig:FMmodel} (c-f) show representative results of the spin densities and the EC from this model. The in-plane spin components have a nonzero divergence near the center of the unit cell, and become large when the gradient of the confinement potential is significant, which leads to a finite $\chi_1^z$. The out-of-plane spin component is largest near the center of the unit cell, where the Laplacian of the confinement potential is also largest, leading to a finite $\chi_2^z$. The summands of Eqs.~\eqref{eq:chi1z} and \eqref{eq:chi2z} are plotted in Figs.~\ref{fig:FMmodel} (d) and (f), respectively. The bright dots represent the (Bragg) peaks in diffraction experiments and the color coding in log scale suggests fast decay versus increasing crystal momentum. The final results are $\chi_1^z = 0.02685$, $\chi_2^z = -0.03058$ [in the units implied in Eqs.~\eqref{eq:chi1z} and \eqref{eq:chi2z}] for the parameter values listed in the caption of Fig.~\ref{fig:FMmodel}.

\section{AHE induced by magnetic charge}\label{sec:new_AHE_systems}

The EC introduced above not only serves as an indicator of the AHE in known materials but also provides intuitive guidance for the search of new AHE systems with vanishing net magnetization. To give a glimpse of the predictive power of EC, in this section we show two experimentally relevant model examples in which the magnetic charge appears explicitly and leads to the AHE, inspired by the way that $\nabla \phi$ and the magnetic charge density $\rho_m$ cooperatively give rise to finite $\boldsymbol{\chi}_1$.

\subsection{Minimal model of the AHE due to magnetic charge order}\label{subsec:charge_order}

\begin{figure}[ht]
	\centering
	\subfloat[]{\includegraphics[width=1.2 in]{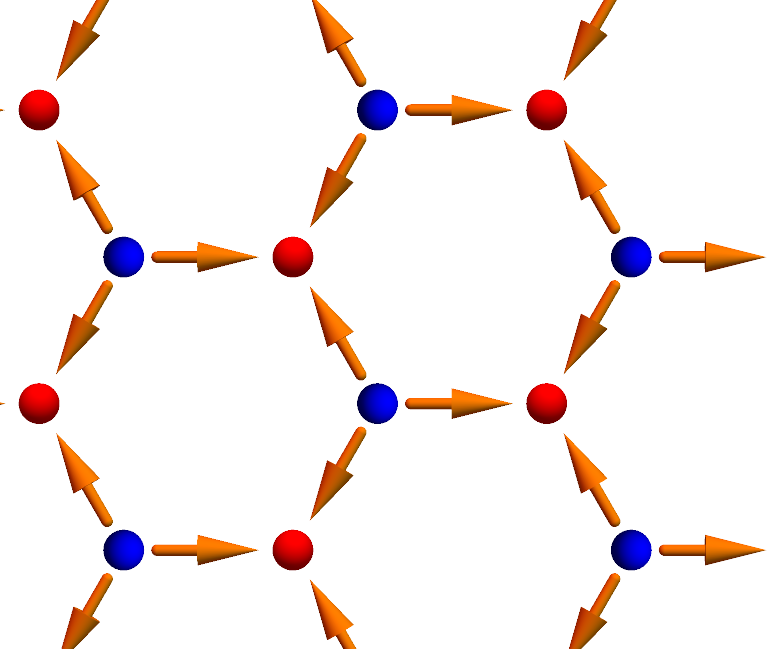}}
	\subfloat[]{\includegraphics[width=1.7 in]{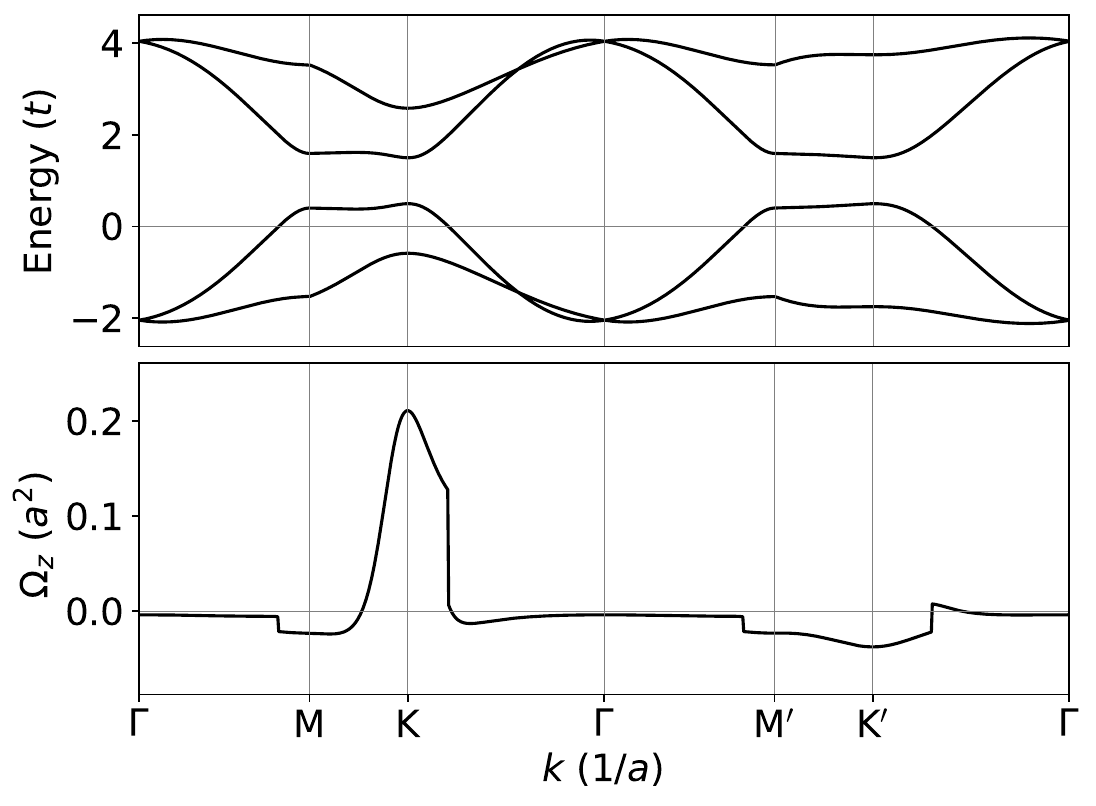}}\\
	\subfloat[]{\includegraphics[width=1.7 in]{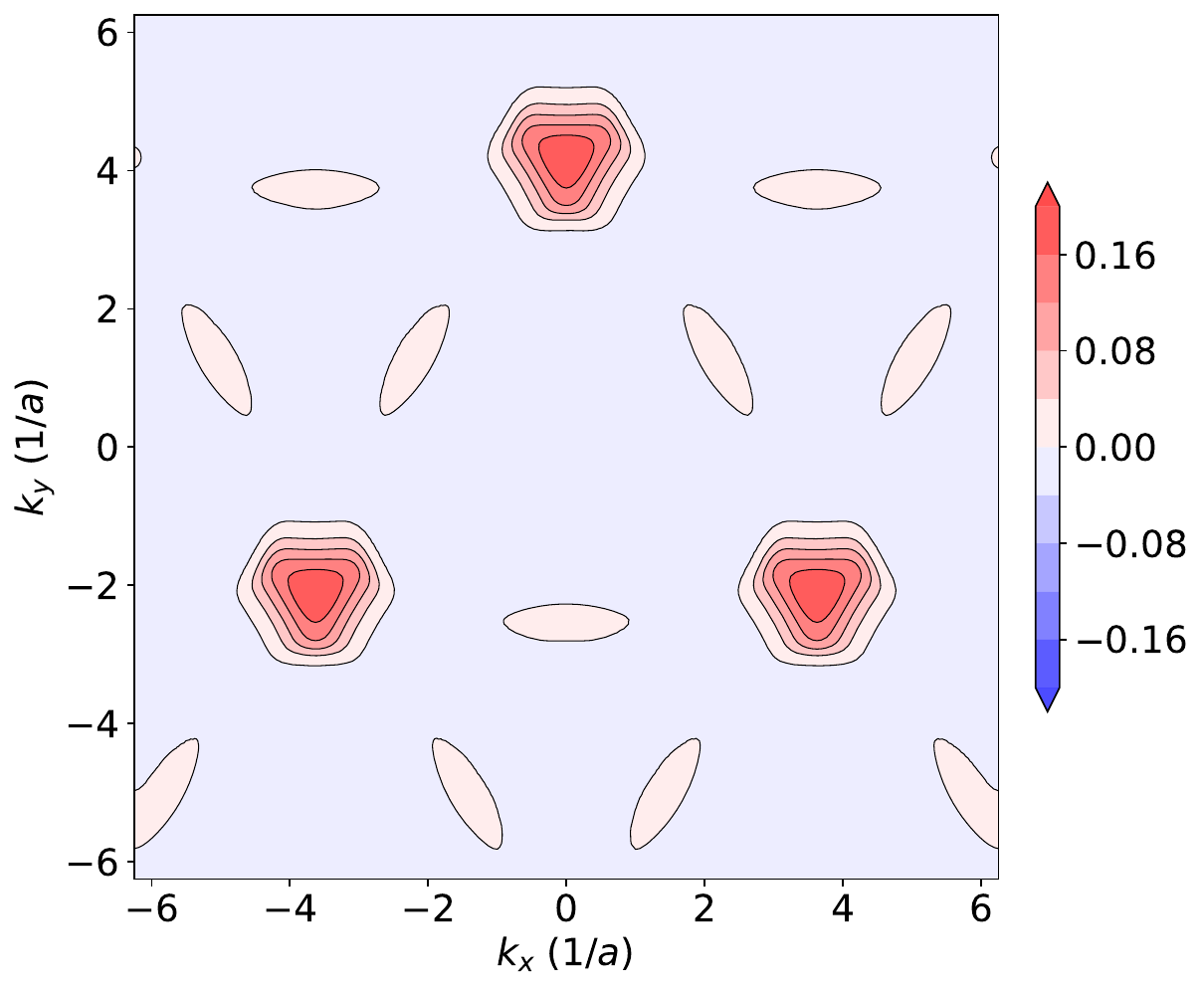}}
	\caption{(Color online) (a) Honeycomb lattice model with opposite magnetic charges residing on the two sublattices, respectively. The arrows correspond to the magnetic field lines. (b) Band structure (top) and Berry curvature summed over occupied bands (bottom) of the model. $\bf M' = -M$ and $\bf K' = - K$. The parameter values are: $t=1, t_M = 0.7, \lambda_R = 0.2, \Delta = 0.5$. (c) Berry curvature obtained using the same parameters as in (b) but plotted in the 2D momentum space.} 
	\label{fig:model}
\end{figure}

We first consider a minimal tight-binding model having the essential ingredients for magnetic-charge-induced AHE. The model describes $s$ electrons hopping between nearest neighbors on a honeycomb lattice, with magnetic charge of opposite signs residing on the two sublattices [Fig.~\ref{fig:model} (a)]:
\begin{eqnarray}\label{eq:hcmodel}
H &=& - t\sum_{\langle i j \rangle \alpha} c_{i\alpha}^\dag c_{j \alpha} - t_M \sum_{\langle i j \rangle \alpha \beta}  \eta_{ij} \boldsymbol{\sigma}_{\alpha\beta} \cdot\hat{r}_{ij} c_{i\alpha}^\dag c_{j \beta} \\\nonumber
&& + \imath \lambda_R \sum_{\langle i j \rangle \alpha \beta} \boldsymbol{\sigma}_{\alpha\beta} \cdot (\hat{z}\times \hat{r}_{ij}) c_{i\alpha}^\dag c_{j \beta} + \Delta \sum_{i\alpha} \gamma_i   c_{i\alpha}^\dag c_{i \alpha}
\end{eqnarray}
where the four terms respectively correspond to spin-independent hopping, spin-dependent hopping due to the magnetic charge, Rashba spin-orbit coupling, and an on-site potential breaking the sublattice symmetry; $\eta_{ij} = +1 (-1)$ if $\hat{r}_{ij}$ points from sublattice A (B) to B (A); $\eta_{ij}$ together with $\boldsymbol{\sigma}\cdot \hat{r}_{ij}$ capture the spin-dependent hopping due to the magnetic field ($\mathbf H$ field) lines between neighboring magnetic charges; $\gamma_i = +1 (-1)$ on A (B) sublattice. The Rashba term is needed to provide the direction of $\boldsymbol{\chi}_1$ along $z$ as suggested by the expression of $\boldsymbol{\chi}_1$, and the sublattice potential is needed to break the degeneracy between the opposite magnetic charges on the two sublattices.

Figure~\ref{fig:model} (b) shows the typical band structure and Berry curvature of model \eqref{eq:hcmodel}. The Berry curvature is generally nonzero and is largest near the Brillouin zone corners $K,K'$. However, the Berry curvatures at opposite momenta do not cancel each other due to the broken time-reversal symmetry by the $t_M$ term, which is also evident from the 2D plot in Fig.~\ref{fig:model} (c). Therefore the AHE is generally nonzero, except when the Fermi energy is in the gap opened by $\Delta$ which we explain next.

After Fourier transform the toy model becomes [for convenience we rotate the honeycomb lattice in Fig.~\ref{fig:model} (a) clockwisely by $\pi/6$]
\begin{widetext}
\begin{eqnarray}
	H &=& \sum_{\mathbf k} C_{\mathbf k}^\dag \begin{pmatrix}
		\Delta \sigma_0 & -t f_{\mathbf k}\sigma_0  - [t_M \mathbf g_{\mathbf k} - \imath \lambda_R (\hat{z}\times \mathbf g_{\mathbf k})]\cdot \boldsymbol{\sigma}\\ 
		-t f^*_{\mathbf k}\sigma_0  - [t_M \mathbf g^*_{\mathbf k} + \imath \lambda_R (\hat{z}\times \mathbf g^*_{\mathbf k})]\cdot \boldsymbol{\sigma} & - \Delta \sigma_0
	\end{pmatrix} C_{\mathbf k} \\\nonumber
	&\equiv& \sum_{\mathbf k} C_{\mathbf k}^\dag h(\mathbf k)  C_{\mathbf k}
\end{eqnarray}
\end{widetext}
where $ C_{\mathbf k}^\dag = (c_{\mathbf k A \uparrow }^\dag, c_{\mathbf k A \downarrow }^\dag, c_{\mathbf k B \uparrow }^\dag, c_{\mathbf k B \downarrow }^\dag)$,
\begin{eqnarray}
	f_{\mathbf k} &=& \sum_{n=1}^3 e^{\imath \frac{a}{\sqrt{3}}\mathbf k\cdot \mathbf r_n} \\\nonumber
	\mathbf g_{\mathbf k} &=& \sum_{n=1}^3 \mathbf r_n e^{\imath \frac{a}{\sqrt{3}}\mathbf k\cdot \mathbf r_n},
\end{eqnarray}
and $\mathbf r_1 =\frac{\sqrt{3}}{2}\hat{x} - \frac{1}{2} \hat{y}, \mathbf r_2 = \hat{y}, \mathbf r_3 =- \frac{\sqrt{3}}{2}\hat{x} - \frac{1}{2} \hat{y}$. $a$ is the lattice constant. Near $\pm \mathbf K = \pm \frac{4\pi}{3 a}\hat{x} $ we have
\begin{eqnarray}
	f_{\mathbf q \pm \mathbf K} &\approx& -\frac{\sqrt{3}a}{2} (\pm q_x - \imath  q_y) + O(q^2),\\\nonumber
	\mathbf g_{\mathbf q \pm \mathbf K} &\approx& \frac{3}{2}(\pm \imath  \hat{x} + \hat{y}) + O(q).
\end{eqnarray}
The Dirac Hamiltonian at each valley $\pm \mathbf K \equiv \eta_v \mathbf K$ is therefore
\begin{flalign}
	& h(\mathbf q + \eta_v \mathbf K) \approx \Delta \tau_z \sigma_0 + \frac{\sqrt{3}at}{2} (\eta_v q_x \tau_x + q_y \tau_y)\sigma_0 & \\\nonumber
	&-\frac{3}{2} (t_M + \eta_v \lambda_R) (\tau_x\sigma_y - \eta_v \tau_y \sigma_x),&
\end{flalign}
where $\boldsymbol{\tau}$ is the Pauli matrix vector in the sublattice space. One can see that the magnetic-charge contribution only changes the magnitude of the Rashba terms in each valley. Since the Rashba terms do not gap the Dirac Hamiltonian by themselves \cite{tse_2011}, the magnetic charge contribution cannot lead to topological phase transitions. Separately, since $t_M$ effectively changes the relative strength of the Rashba spin-orbit coupling at the two valleys, when the Fermi energy is not within the gap, the Berry curvature at the two valleys will not cancel out, leading to the finite anomalous Hall conductivity.

The minimal model can be connected with the magnetic-charge-ordered state of the kagome spin ice \cite{moller_2009,chern_2011,chern_2012} by the duality between honeycomb and kagome lattices. The background magnetic field connecting neighboring magnetic charges can be regarded, as a first approximation, as the homogenized effect of the fluctuating magnetic dipole moments in the charge-ordered state of the kagome spin ice on itinerant electrons. In comparison with the models studied in \cite{ishizuka_2013,chern_2014} where the local spins on the kagome lattice is noncoplanar, the present model has a vanishing net magnetization. More importantly, the essential symmetry breaking in the non-magnetic part of model \eqref{eq:hcmodel} is already present in the pyrochlore iridate Pr$_2$Ir$_2$O$_7$ \cite{Tomizawa_2009}, in which an AHE in the absence of long-range dipolar order and of net magnetization has been observed \cite{Machida_AHE_no_order_2010}. Although the ground state of Pr$_2$Ir$_2$O$_7$ may be elusive and the direct measurement \cite{vanderLaan_2021} of scalar spin chirality by scattering techniques is challenging, it is possible to alternatively measure the EC which, if nonzero, can help to solve the puzzle of the zero-field AHE in \cite{Machida_AHE_no_order_2010}. 

\subsection{Skew scattering by magnetic charge}\label{subsec:skewscattering}

In this subsection we predict an extrinsic contribution to the AHE by magnetic charge through skew scattering. Again motivated by the expression of $\boldsymbol{\chi}_1$, we consider the following model of 2D Dirac electrons with Rashba-type spin-momentum locking scattered by a magnetic charge whose magnetic field is truncated at finite radius $R$:
\begin{eqnarray}\label{eq:H}
H &=& - \imath \hbar \lambda (\sigma_x\partial_y - \sigma_y \partial_x) - \frac{\Delta}{2\pi r}\hat{r}\cdot \boldsymbol{\sigma}\Theta(R-r) \\\nonumber
&\equiv& H_D + H_\Delta
\end{eqnarray} 
where $\Theta(x)$ is the step function. $H_\Delta$ represents the Zeeman coupling between the electron spin and the magnetic field $\mathbf h(\mathbf r) = - \mathbf m(\mathbf r) = \Delta\hat{r}/(2\pi r)$ generated by a magnetic charge located at the origin within a radius $R$. Here we consider the analytically simpler case of $\alpha \equiv \Delta/(2\pi\hbar\lambda) = 1$ and relegate the more general solution to Appendix~\ref{appendix:skew}. Assuming a positive chemical potential, the solution for $r<R$ with energy $E = \hbar \lambda k_0 > 0$ is 
\begin{eqnarray}
\Psi_{<} = \sum_{n=-\infty}^\infty a_n \begin{pmatrix}
J_{n-1}(k_0 r)e^{{\imath } n \theta} \\
J_{n}(k_0 r)e^{{\imath } (n+1) \theta}
\end{pmatrix}
\end{eqnarray}
while that for $r>R$ with an incident plane wave traveling along $\hat{x}$ is
\begin{flalign}
&\Psi_{>} = \Psi_{\rm in} + \Psi_{\rm scatt} &\\\nonumber
&=\frac{e^{\imath k_0x}}{2\sqrt{2}\pi} \begin{pmatrix}
\imath \\
1
\end{pmatrix} + \sum_{n = -\infty}^{\infty}b_n \begin{pmatrix}
H_{n}(k_0 r)e^{\imath n\theta} \\
H_{n+1}(k_0 r)e^{\imath (n+1)\theta}
\end{pmatrix}&
\end{flalign}
where $H_n = J_n + \imath Y_n$ is the Hankel function of the first kind, and $Y_n$ is the Bessel function of the second kind. The appearance of the Hankel function is because $\Psi_>$ does not include the origin where $Y_n$ diverges. Also the Hankel function of the first kind represents outgoing waves \cite{Smit_1958}. Solving $b_n$ from the boundary condition $\Psi_<|_{r=R} = \Psi_>|_{r=R}$ and taking the large-distance asymptotic form of $\Psi_{\rm scatt}$, we obtain the scattering cross section
\begin{eqnarray}
	\sigma(\theta)\propto \langle \mathbf j\cdot \hat{r} \rangle_{\rm scatt} (r,\theta) = \frac{4\lambda}{\pi k_0 r}\Big | \sum_{n} b_n e^{{\imath}n(\theta-\frac{\pi}{2})} \Big |^2.
\end{eqnarray}
When $k_0 R \ll 1$ one can consider up to $p$-wave contributions to $\sigma (\theta)$. The Hall angle due to scattering by the magnetic charge only can be calculated as \cite{ferreira_2014}
\begin{eqnarray}
	\tan\theta_{\rm H} &=& \frac{\int \sigma(\theta)\sin\theta d\theta}{\int \sigma(\theta)(1-\cos\theta)d\theta} \approx {\rm Re}\left(\frac{b_{-1} + b_1}{b_0}\right)\\\nonumber
	&\approx& \frac{\pi}{8} (k_0 R)^2
\end{eqnarray}
The result in the last line above turns out to be a good approximation even at $k_0 R \sim 1$, for which $\tan\theta_{\rm H}\approx 0.39$. We also found that when $\alpha = -1$, $\tan\theta_{\rm H} \approx -\frac{\pi}{8} (k_0 R)^2$, confirming the time-reversal-odd property of the AHE. 

Experimental detection of such an effect may be performed using topological insulator surface states \cite{Hsieh_2009} or 2D electron gas with large Rashba spin-orbit coupling that are proximate-coupled to magnetic textures \cite{araki_2017,ishizuka_2018} having a nonzero 2D magnetic charge density. With $\hbar \lambda \sim 10^1\, {\rm eV \cdot \AA}$ and $R\sim 1/k_0 \sim 100$ \AA, the Zeeman coupling $\Delta/(2\pi R)\sim 0.1$ eV is reasonable to achieve experimentally. 

\section{Discussion}\label{sec:discussion}
 
The electronic chiralization introduced in this work is a construction based on charge and spin densities that themselves are physical observables measured by scattering techniques. Therefore it does not directly correspond to a thermodynamic variable that is conjugate to a single external field configuration, such as the magnetization or toroidization. However, a corresponding thermodynamic variable for EC may be defined through the coupling with multipole moments of non-Gaussian electromagnetic waves. First-principles calculations of the EC in plane-wave basis is also straightforward, as demonstrated in Sec.~\ref{subsec:EC_FMRashba}, but to get meaningful values a proper treatment of the cutoff (e.g., by using an atomic form factor) may be essential due to the fluctuation of the gradients of charge and spin densities at high energies. Alternatively, one may use the expressions of EC derived by assuming localized atomic magnetic moments and charge.

Although we mainly focused on the magnetic-charge-related $\boldsymbol{\chi}_1$, it is possible to predict the existence of the AHE in other systems based on the forms of $\boldsymbol{\chi}_{2,3}$ and their generalizations. For example, the use of a sublattice potential in the honeycomb model in Sec.~\ref{subsec:charge_order} makes the AHE more precisely correspond to $\boldsymbol{\chi}''_1$ in Eq.~\eqref{eq:EChigher}. We stress that such predictions based on the grounds of symmetry do not necessarily yield universal microscopic mechanisms for the AHE, since similar to ferromagnets, in a given system with nonzero EC all mechanisms relevant to the AHE should generally coexist if without fine-tuning. Nonetheless, the examples given in Sec.~\ref{sec:new_AHE_systems} suggest that there are new “building blocks” for the intrinsic and extrinsic mechanisms of the AHE inspired by the EC, similar to the case of scalar spin chirality \cite{Tokura_2018,ishizuka_2018}. Finally, EC can be used as an indicator of other anomalous response functions such as the anomalous Nernst effect or magneto-optical Kerr effect that have similar symmetry properties as the AHE. 

\begin{acknowledgements} 
H.C. was supported by NSF CAREER Grant No. DMR-1945023. H.C. is grateful to A. MacDonald, O. Pinaud, K. Zhao, and K. Ross for valuable discussions.
\end{acknowledgements}

\appendix
\section{Real-space formulas for the generalized EC}\label{appendix:gEC}
Real-space formulas for the generalized EC constructions in Eq.~\eqref{eq:EChigher} can be generated by integrals of three or more Gaussians:
\begin{flalign}\label{eq:Ian}
	&I(\mathbf a_1, \mathbf a_2, \dots, \mathbf a_n) = \int d^3\mathbf r \prod_{i=1}^n g(\mathbf r -\mathbf a_i)&\\\nonumber
	&= (2\pi\sigma^2)^{-\frac{3(n-1)}{2}}n^{-\frac{3}{2}} \exp\left[ -\frac{n}{2\sigma^2}\left( \langle a^2\rangle -\langle \mathbf a\rangle \cdot \langle \mathbf a \rangle   \right)  \right]&
\end{flalign}
where
\begin{eqnarray}
	\langle a^2\rangle =\frac{1}{n}\sum_{i=1}^n a_i^2,\;\langle \mathbf a \rangle =\frac{1}{n}\sum_{i=1}^n \mathbf a_i
\end{eqnarray}
Using Eq.~\eqref{eq:Ian} we can obtain
\begin{flalign}
	&	I_{ij} (\mathbf a, \mathbf b,\mathbf c) \equiv \int d^3\mathbf r g(\mathbf r - \mathbf a) \partial_ig(\mathbf r -\mathbf b) \partial_j g(\mathbf r -\mathbf c)& \\\nonumber
	&= \partial_{b_i} \partial_{c_j} I(\mathbf a, \mathbf b, \mathbf c)&\\\nonumber
	&=\frac{1}{24\pi^3 3^{\frac{3}{2}} \sigma^8} \left[\delta_{ij} + \frac{1}{3\sigma^2} (2b_i - a_i - c_i) (2c_j - a_j - b_j) \right]&\\\nonumber
	&\times \exp\left[-\frac{1}{3\sigma^2} (a^2+b^2+c^2 - \mathbf a \cdot \mathbf b - \mathbf b \cdot \mathbf c - \mathbf c \cdot \mathbf a) \right]&
\end{flalign}
which enters the real-space expression of $\boldsymbol{\chi}'_1$
\begin{eqnarray}
	\chi'_{1i} = \frac{1}{V_{\rm uc}} \sum_{\mathbf R\mathbf R'} \sum_{mnp}  Q_m Q_n M^j_p I_{ij}(\mathbf r_m , \mathbf r_n + \mathbf R, \mathbf r_p + \mathbf R')
\end{eqnarray}
Because the exponent in $I_{ij}$ is proportional to the standard deviation of the positions of the three sites in the summand, only near neighbors need to be considered.

For the calculation of $\boldsymbol{\chi}_1''$ we need
\begin{eqnarray}
	I_{ijkl} (\mathbf a,\mathbf b, \mathbf c) \equiv \partial_{a_i}\partial_{b_j}\partial_{b_k} \partial_{c_l} I(\mathbf a, \mathbf b, \mathbf c).
\end{eqnarray}
The evaluation of this quantity can be simplified by defining the following Feynman rules. Writing the exponent in $I$ as $f$, one can see that derivatives of $f$ higher than 2nd order will vanish. One can therefore represent the variables in the derivatives as nodes and $f$ as lines. 1st and 2nd derivatives of $f$ can be represented by an open-ended line and a line connecting two nodes, respectively. $I_{ijkl}$ can therefore be represented by a sum over topologically distinct diagrams of 4 nodes. There are in total 10 diagrams: 1 with 4 lines, 6 with 3 lines, and 3 with 2 lines. If $j=k$, there are 7 terms in the expression of $I_{ijjl}$
\begin{widetext}
	\begin{eqnarray}\label{eq:Iijjl}
		I_{ijjl} &=& \frac{e^f}{8\pi^3 3^{\frac{3}{2}} \sigma^6} \big[ 2 (\partial_{a_i}\partial_{b_j} f)( \partial_{b_j}\partial_{c_l} f) + (\partial_{a_i}\partial_{c_l} f) (\partial^2_{b_j} f) + 2 (\partial_{a_i}\partial_{b_j} f) (\partial_{b_j} f) (\partial_{c_l} f) \\\nonumber
		&+&  (\partial_{a_i}\partial_{c_l} f) (\partial_{b_j}f)^2 + (\partial_{a_i}f) (\partial^2_{b_j} f) (\partial_{c_l} f) + 2 (\partial_{a_i}f) (\partial_{b_j} f) (\partial_{b_j}\partial_{c_l} f) + (\partial_{a_i} f) (\partial_{b_j} f) (\partial_{b_j} f) (\partial_{c_l} f)\big]\\\nonumber
		&=& \frac{e^f}{8\pi^3 3^{\frac{3}{2}} \sigma^6}\Big[-\frac{4}{9\sigma^4} \delta_{il} - \frac{2}{27\sigma^6} (2b_i - a_i - c_i)(2c_l - a_l - b_l) -  \frac{1}{27\sigma^6} \delta_{il} |2\mathbf b -\mathbf a - \mathbf c|^2 \\\nonumber
		&+&  \frac{2}{9\sigma^6} (2a_i - b_i - c_i)(2c_l - a_l - b_l) - \frac{2}{27\sigma^6} (2a_i - b_i - c_i) (2b_l - a_l - c_l) \\\nonumber
		&+& \frac{1}{81\sigma^8} (2a_i - b_i - c_i)(2c_l - a_l - b_l) |2\mathbf b -\mathbf a - \mathbf c|^2 \Big] \\\nonumber
		&\equiv& \frac{e^f}{72\pi^3 3^{\frac{3}{2}} \sigma^{10}} \Big[- \left(4 +  \frac{|\tilde{\mathbf b}|^2}{3\sigma^2} \right)\mathbb{I} -\frac{2}{3\sigma^2} (\tilde{\mathbf b}\tilde{\mathbf c} +\tilde{\mathbf a}\tilde{\mathbf b}) + \left( \frac{2}{\sigma^2} + \frac{|\tilde{\mathbf b}|^2}{9\sigma^4} \right)\tilde{\mathbf a}\tilde{\mathbf c} \Big]_{il} 
	\end{eqnarray}
\end{widetext}
For Fe$_2$O$_3$ the cluster with two Fe atoms sandwiched between three layers of O has $D_3$ symmetry \cite{DZYALOSHINSKY_1958,moriya_1960}. The EC would have been forbidden if the symmetry were $D_{3h}$, i.e., if the top and bottom oxygen layers were not distorted. The nonzero contribution to $\boldsymbol{\chi}_1''$ comes from the last term in Eq.~\eqref{eq:Iijjl}.

\section{Dirac electrons scattered by a magnetic charge}\label{appendix:skew}
Ignoring the step function $\Theta(R - r)$ first, the Hamiltonian in Eq.~\eqref{eq:H} can be written in the polar coordinates as
\begin{eqnarray}
	H = \begin{pmatrix}
		0 &  f\\
		f^\dag &  0
	\end{pmatrix}.
\end{eqnarray}
where $f\equiv e^{-\imath \theta} \left[ \hbar\lambda \left( \partial_r - \frac{\imath}{r} \partial_\theta   \right)  -\frac{\Delta}{2\pi r} \right]$. Since the Hamiltonian is invariant under rotation with respect to the $z$ axis going through the origin, the total angular momentum along $z$ is a good quantum number:
\begin{eqnarray}
	J_z &=& L_z + s_z = -\imath \hbar \partial_\theta + \frac{\hbar}{2}\sigma_z.
\end{eqnarray}
$J_z$ satisfies the following eigenequation
\begin{eqnarray}
	J_z \begin{pmatrix}
		e^{\imath n\theta} \\
		e^{\imath (n+1)\theta}
	\end{pmatrix} = \left( n + \frac{1}{2}\right) \hbar \begin{pmatrix}
		e^{\imath n\theta} \\
		e^{\imath (n+1)\theta}
	\end{pmatrix} 
\end{eqnarray}
Therefore we can take the following trial solution
\begin{eqnarray}
	\psi = \begin{pmatrix}
		u(r) e^{\imath n\theta} \\
		v(r) e^{\imath (n+1)\theta}
	\end{pmatrix}.
\end{eqnarray}
The resulting radial equations are
\begin{eqnarray}\label{eq:HRpHD}
	&& \frac{E}{\hbar\lambda} u - \left( \partial_r + \frac{n+1}{r}  \right) v + \frac{\alpha}{r} v = 0,  \\\nonumber
	&& \frac{E}{\hbar \lambda} v - \left( -\partial_r + \frac{n}{r}  \right) u + \frac{\alpha}{r} u  = 0,  
\end{eqnarray}
where $\alpha \equiv \Delta/(2\pi\hbar\lambda)$. 

When $\alpha$ is an integer, using the recurrence relations of the Bessel functions one can immediately see $u\propto J_{n-\alpha}(\kappa r)$, $v\propto J_{n-\alpha+1}(\kappa r)$, and $E=\pm \hbar \lambda \kappa$. The normalized eigenfunction is
\begin{eqnarray}\label{eq:solHrHd}
	\psi_{p,n,\kappa} = \sqrt{\frac{\kappa}{4\pi}}\begin{pmatrix}
		J_{n-\alpha}(\kappa r) e^{\imath n\theta} \\ 
		p J_{n-\alpha +1} (\kappa r) e^{\imath (n+1) \theta}
	\end{pmatrix}
\end{eqnarray}
where $p=\pm$. Note that when $\alpha$ is an integer, the $H_\Delta$ term can be removed by a gauge transformation:
\begin{eqnarray}
	\psi\rightarrow \psi e^{\imath \alpha\theta},
\end{eqnarray}
which is the reason why $\psi_{p,n,\kappa}$ becomes an eigenstate ($\psi_{p,n-\alpha,\kappa}$) of $H_D$ in this case. The magnetic charge is nonetheless still able to induce skew scattering because the truncation $\Theta(R - r)$ makes it impossible for the $H_\Delta$ term to be removed by a pure gauge transformation.

When $\alpha$ is not an integer the solution is less trivial. To avoid ambiguity we consider the magnetic charge potential regularized by replacing $\Delta$ with $\Delta\Theta(r-r_0)$. The solution for $r<r_0$ is simply (the arguments of the Bessel functions are omitted for brevity)
\begin{eqnarray}\label{eq:eigvecR}
	\psi_< = \sqrt{\frac{\kappa}{4\pi}}\begin{pmatrix}
		J_{n} e^{\imath n\theta} \\ 
		p J_{n +1} e^{\imath (n+1) \theta}
	\end{pmatrix}
\end{eqnarray}
while that for $r>r_0$ involves both $J_\nu$ and $J_{-\nu}$ which are linearly independent solutions of the Bessel equation when $\nu$ is not an integer. More explicitly:
\begin{eqnarray}\label{eq:solreg}
	\psi_> =\sqrt{\frac{\kappa}{4\pi}} \begin{pmatrix}
		(A J_{n-\alpha} + B J_{-n + \alpha}) e^{\imath n\theta} \\ 
		p (AJ_{n-\alpha +1} - BJ_{-n+\alpha -1}) e^{\imath (n+1) \theta}
	\end{pmatrix}
\end{eqnarray}
where $A,B$ are coefficients depending on $n$.

Using the boundary condition $\psi_<(r=r_0) = \psi_>(r=r_0)$ leads to 
\begin{eqnarray}
	A &=& \frac{J_{-n+\alpha-1} J_{n} + J_{-n+\alpha} J_{n+1} }{J_{-n+\alpha} J_{n-\alpha +1} + J_{-n+\alpha -1} J_{n-\alpha}} \\\nonumber
	B &=& \frac{J_{n-\alpha+1} J_{n} - J_{n-\alpha} J_{n+1} }{J_{-n+\alpha} J_{n-\alpha +1} + J_{-n+\alpha -1} J_{n-\alpha}}
\end{eqnarray}
where the arguments of the Bessel functions are all $\kappa r_0$. To understand the asymptotic behavior as $r_0 \rightarrow 0$ we consider the ratio $A/B$:
\begin{eqnarray}
	\frac{A}{B} = \frac{J_{-n+\alpha-1} J_{n} + J_{-n+\alpha} J_{n+1} }{J_{n-\alpha+1} J_{n} - J_{n-\alpha} J_{n+1}}.
\end{eqnarray}
We found that when $\alpha > 0$, $A/B$ diverges when $n>\alpha-1$ and vanishes when $n<\alpha - 1$. However, when $\alpha < 0$, $A/B$ diverges when $n>\alpha$ and vanishes when $n<\alpha$. Therefore the solution is, for $\alpha > 0$:
\begin{flalign}\label{eq:solHRpHDfull1}
	&\psi_{p,n,\kappa} = &\\\nonumber
	&\sqrt{\frac{\kappa}{4\pi}} \times \begin{cases}
		\begin{pmatrix}
			J_{n-\alpha}(\kappa r) e^{\imath n\theta} \\ 
			p J_{n-\alpha +1} (\kappa r) e^{\imath (n+1) \theta}
		\end{pmatrix} & n > \alpha - 1,\\
		\begin{pmatrix}
			J_{-n+\alpha}(\kappa r) e^{\imath n\theta} \\ 
			-p J_{-n+\alpha -1} (\kappa r) e^{\imath (n+1) \theta}
		\end{pmatrix} & n < \alpha - 1 ,
	\end{cases}&
\end{flalign}
and for $\alpha < 0$:
\begin{flalign}\label{eq:solHRpHDfull2}
	&\psi_{p,n,\kappa} = &\\\nonumber
	&\sqrt{\frac{\kappa}{4\pi}} \times \begin{cases}
		\begin{pmatrix}
			J_{n-\alpha}(\kappa r) e^{\imath n\theta} \\ 
			p J_{n-\alpha +1} (\kappa r) e^{\imath (n+1) \theta}
		\end{pmatrix} & n > \alpha,\\
		\begin{pmatrix}
			J_{-n+\alpha}(\kappa r) e^{\imath n\theta} \\ 
			-p J_{-n+\alpha -1} (\kappa r) e^{\imath (n+1) \theta}
		\end{pmatrix} & n < \alpha
	\end{cases}&
\end{flalign}
One can check that when $\alpha = 0$, the above solution returns to Eq.~\eqref{eq:eigvecR}. In addition, when $\alpha$ is an arbitrary integer, by using the relation
\begin{eqnarray}
	J_{-n}(x) = (-1)^n J_n (x)
\end{eqnarray}
Eq.~\eqref{eq:solHrHd} is recovered.

We next calculate the skew scattering by considering the truncated magnetic charge potential:
\begin{eqnarray}
	H_{\Delta }= - \frac{\Delta}{2\pi r}\hat{r}\cdot \boldsymbol{\sigma}\Theta(R-r).
\end{eqnarray}
The solution for $r<R$ with energy $\hbar \lambda k_0$ is 
\begin{eqnarray}
	\Psi_{<} = \sum_{n=-\infty}^\infty a_n  \sqrt{\frac{4\pi}{k_0}}  \psi_{+,n,k_0}
\end{eqnarray}
where $\psi_{+,n,k_0}$ is given in Eqs.~\eqref{eq:solHrHd}, \eqref{eq:solHRpHDfull1}, or \eqref{eq:solHRpHDfull2} depending on the value of $\alpha$, while that for $r>R$ is  
\begin{flalign}
&	\Psi_{>} = \sum_{n=-\infty}^{\infty} \frac{1}{2\sqrt{2}\pi} \imath ^n J_n(k_0 r) e^{\imath n \theta} \begin{pmatrix}
		\imath \\
		1
	\end{pmatrix}&\\\nonumber
& + \sum_{n = -\infty}^{\infty}b_n \begin{pmatrix}
		H_{n}(k_0 r)e^{\imath n\theta} \\
		H_{n+1}(k_0 r)e^{\imath (n+1)\theta}
	\end{pmatrix}.&
\end{flalign}
where $H_n = J_n + \imath Y_n$ is the Hankel function of the first kind \cite{Smit_1958}, and $Y_n$ is the Bessel function of the second kind. The boundary condition $\Psi_<(r=R) = \Psi_>(r=R)$ leads to the following relation for $n>\alpha-1$ ($\alpha > 0$) or $n>\alpha$ ($\alpha<0$):
\begin{eqnarray}\label{eq:bceqn}
	&&a_n J_{n-\alpha} - b_n H_n = \frac{\imath ^{n+1}}{2\sqrt{2}\pi} J_{n} \\\nonumber
	&&a_n J_{n-\alpha+1} - b_n H_{n+1} = \frac{\imath ^{n+1}}{2\sqrt{2}\pi} J_{n+1}
\end{eqnarray}
which has the solution
\begin{flalign}\label{eq:anbnsol}
&	\begin{pmatrix}
		a_n \\ 
		b_n
	\end{pmatrix} = \frac{\imath ^{n+1}}{2\sqrt{2}\pi}\frac{1}{H_n J_{n-\alpha + 1} - H_{n+1} J_{n-\alpha}}  &\\\nonumber
& \times\begin{pmatrix}
		H_n J_{n+1} - H_{n+1} J_n \\
		J_{n+1} J_{n-\alpha} - J_n J_{n-\alpha + 1}
	\end{pmatrix}.&
\end{flalign}
The above result also applies to any $n$ when $\alpha$ is an integer. 

When $n<\alpha-1$ ($\alpha > 0$) or $n<\alpha$ ($\alpha<0$), we have
\begin{flalign}\label{eq:anbnsoll0}
	&\begin{pmatrix}
		a_n \\ 
		b_n
	\end{pmatrix} = -\frac{\imath ^{n+1}}{2\sqrt{2}\pi}\frac{1}{H_n J_{-n+\alpha - 1} + H_{n+1} J_{-n+\alpha}} &\\\nonumber
&\times \begin{pmatrix}
		H_n J_{n+1} - H_{n+1} J_n \\
		J_{n+1} J_{-n+\alpha} + J_n J_{-n+\alpha - 1}
	\end{pmatrix}.&
\end{flalign}

To go further, we consider the limit $k_0 r\gg 1$ and use the asymptotic form of $H_\nu$:
\begin{eqnarray}
	H_\nu(x) \approx \sqrt{\frac{2}{\pi x}} e^{\imath \left(x-\frac{\nu\pi}{2}-\frac{\pi}{4} \right)},
\end{eqnarray}
so that the scattered wave becomes 
\begin{flalign}\label{eq:psiscatt}
	&\Psi_{\rm scatt} = \Psi_{>} - \Psi_{\rm in}& \\\nonumber
	&\approx \sum_{n = -\infty}^{\infty}\sqrt{\frac{2}{\pi k_0 r}} b_n\begin{pmatrix}
		e^{\imath \left[k_0 r - \frac{n\pi}{2} - \frac{\pi}{4} +  n\theta\right]} \\
		e^{\imath \left[k_0 r - \frac{(n+1)\pi}{2} - \frac{\pi}{4} + (n+1)\theta\right]}&
	\end{pmatrix}.
\end{flalign}

The scattering cross section is
\begin{flalign}
&	\sigma(\theta)=&\\\nonumber
&\frac{ \Psi_{\rm scatt}^\dag (\mathbf j \cdot \mathbf r) \Psi_{\rm scatt} }{\Psi_{\rm in}^\dag j_x \Psi_{\rm in}} =  \frac{16\pi}{k_0} \Big| \sum_n b_n e^{\imath n(\theta -\frac{\pi}{2})}\Big |^2.&
\end{flalign}
The skew cross section is therefore
\begin{flalign}
	&\sigma_{\rm skew} \equiv &\\\nonumber
	&\int_0^{2\pi} \sin\theta \sigma(\theta) d\theta = \frac{32 \pi^2}{k_0} \sum_n {\rm Re}(b_{n+1}^* b_n).&
\end{flalign}

\begin{figure}[ht]
	\centering
	\includegraphics[width=2.2 in]{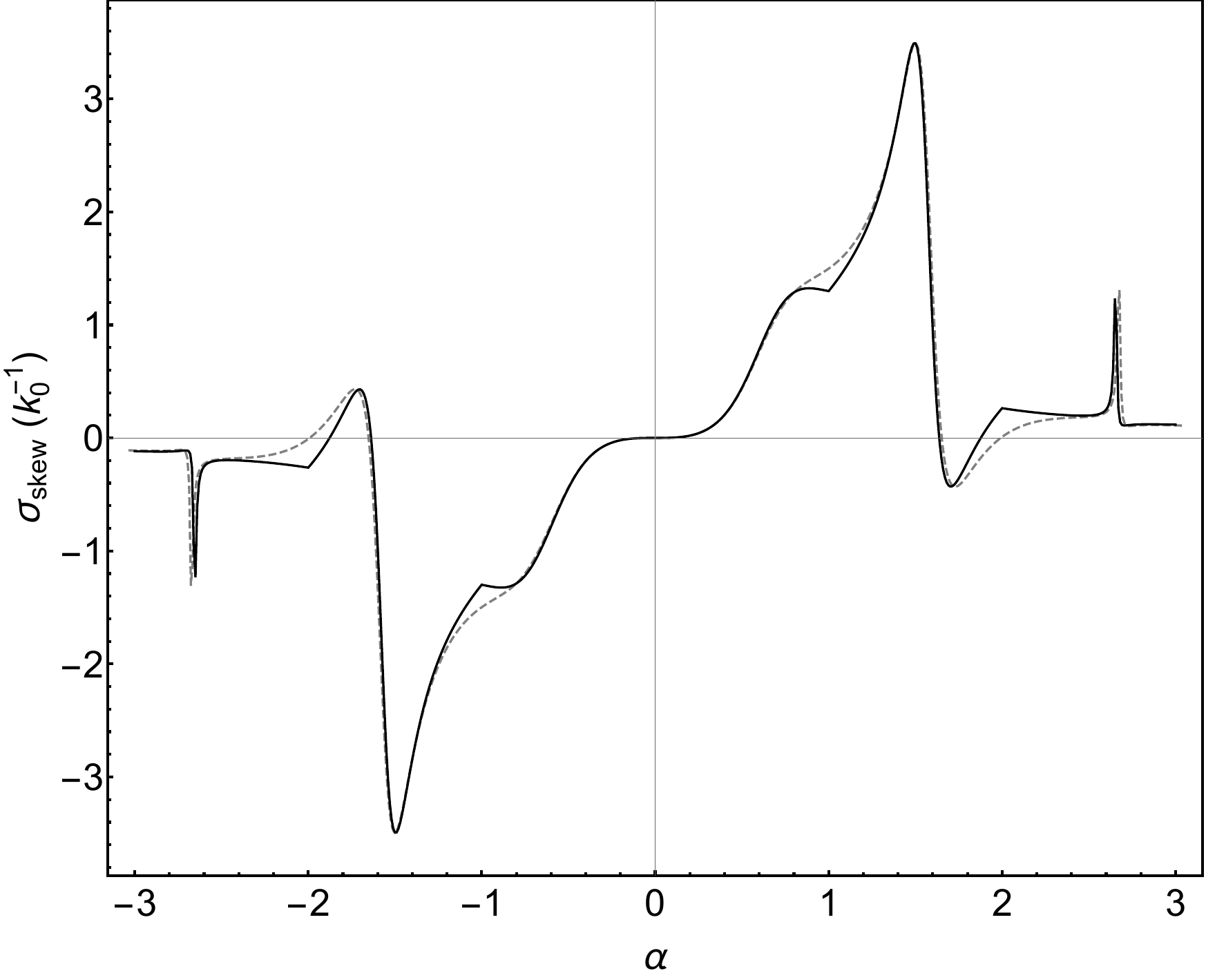}
	\caption{Skew scattering cross section plotted against $\alpha$. $k_0 R = 1$. The solid line is the result by using Eqs.~\eqref{eq:solHRpHDfull1} and \eqref{eq:solHRpHDfull2} as the solution for $r<R$, while the dashed line is the result by using the regularized solution Eq.~\eqref{eq:solreg} for $r<R$, with $k_0 r_0 = 0.01$.} 
	\label{fig:sskew}
\end{figure}

Fig.~\ref{fig:sskew} (solid black line) plots $\sigma_{\rm skew}$ versus $\alpha$. It is evident that $\sigma_{\rm skew}$ is odd under $\alpha\rightarrow -\alpha$. That the skewness decreases quickly with increasing $\alpha>2$ is because the large $\alpha$ effectively suppresses the contributions at small $n$ which give the dominant scattering amplitudes. The apparent discontinuity of $\partial \sigma_{\rm skew} /\partial \alpha $ at integer values of $\alpha\neq 0$ originates from the singularity of $\psi_{p,n,\kappa}$ in Eqs.~\eqref{eq:solHRpHDfull1} and \eqref{eq:solHRpHDfull2} when $\alpha - 1 < n <\alpha$. Such singularity can be removed by regularizing the magnetic charge potential at $r\rightarrow 0$ as in Eq.~\eqref{eq:solreg}. For example, by choosing $k_0r_0 = 0.01$, the discontinuities of the derivative of $\sigma_{\rm skew}$ are absent (gray dashed line in Fig.~\ref{fig:sskew}). 

\bibliography{mcahe_ref}

\begin{thebibliography}{63}%
\makeatletter
\providecommand \@ifxundefined [1]{%
 \@ifx{#1\undefined}
}%
\providecommand \@ifnum [1]{%
 \ifnum #1\expandafter \@firstoftwo
 \else \expandafter \@secondoftwo
 \fi
}%
\providecommand \@ifx [1]{%
 \ifx #1\expandafter \@firstoftwo
 \else \expandafter \@secondoftwo
 \fi
}%
\providecommand \natexlab [1]{#1}%
\providecommand \enquote  [1]{``#1''}%
\providecommand \bibnamefont  [1]{#1}%
\providecommand \bibfnamefont [1]{#1}%
\providecommand \citenamefont [1]{#1}%
\providecommand \href@noop [0]{\@secondoftwo}%
\providecommand \href [0]{\begingroup \@sanitize@url \@href}%
\providecommand \@href[1]{\@@startlink{#1}\@@href}%
\providecommand \@@href[1]{\endgroup#1\@@endlink}%
\providecommand \@sanitize@url [0]{\catcode `\\12\catcode `\$12\catcode
  `\&12\catcode `\#12\catcode `\^12\catcode `\_12\catcode `\%12\relax}%
\providecommand \@@startlink[1]{}%
\providecommand \@@endlink[0]{}%
\providecommand \url  [0]{\begingroup\@sanitize@url \@url }%
\providecommand \@url [1]{\endgroup\@href {#1}{\urlprefix }}%
\providecommand \urlprefix  [0]{URL }%
\providecommand \Eprint [0]{\href }%
\providecommand \doibase [0]{https://doi.org/}%
\providecommand \selectlanguage [0]{\@gobble}%
\providecommand \bibinfo  [0]{\@secondoftwo}%
\providecommand \bibfield  [0]{\@secondoftwo}%
\providecommand \translation [1]{[#1]}%
\providecommand \BibitemOpen [0]{}%
\providecommand \bibitemStop [0]{}%
\providecommand \bibitemNoStop [0]{.\EOS\space}%
\providecommand \EOS [0]{\spacefactor3000\relax}%
\providecommand \BibitemShut  [1]{\csname bibitem#1\endcsname}%
\let\auto@bib@innerbib\@empty
%</preamble>
\bibitem [{\citenamefont {Hall}(1881)}]{Hall_1881}%
  \BibitemOpen
  \bibfield  {author} {\bibinfo {author} {\bibfnamefont {E.}~\bibnamefont
  {Hall}},\ }\bibfield  {title} {\bibinfo {title} {{XVIII. On the “Rotational
  Coefficient” in nickel and cobalt}},\ }\href
  {https://doi.org/10.1080/14786448108627086} {\bibfield  {journal} {\bibinfo
  {journal} {The London, Edinburgh, and Dublin Philosophical Magazine and
  Journal of Science}\ }\textbf {\bibinfo {volume} {12}},\ \bibinfo {pages}
  {157} (\bibinfo {year} {1881})},\ \Eprint
  {https://arxiv.org/abs/https://doi.org/10.1080/14786448108627086}
  {https://doi.org/10.1080/14786448108627086} \BibitemShut {NoStop}%
\bibitem [{\citenamefont {Smith}(1921)}]{Smith_AHE_1921}%
  \BibitemOpen
  \bibfield  {author} {\bibinfo {author} {\bibfnamefont {A.~W.}\ \bibnamefont
  {Smith}},\ }\bibfield  {title} {\bibinfo {title} {{The Hall Effect and the
  Nernst Effect in Magnetic Alloys}},\ }\href
  {https://doi.org/10.1103/PhysRev.17.23} {\bibfield  {journal} {\bibinfo
  {journal} {Phys. Rev.}\ }\textbf {\bibinfo {volume} {17}},\ \bibinfo {pages}
  {23} (\bibinfo {year} {1921})}\BibitemShut {NoStop}%
\bibitem [{\citenamefont {Nagaosa}\ \emph {et~al.}(2010)\citenamefont
  {Nagaosa}, \citenamefont {Sinova}, \citenamefont {Onoda}, \citenamefont
  {MacDonald},\ and\ \citenamefont {Ong}}]{Nagaosa_AHE_RMP_2010}%
  \BibitemOpen
  \bibfield  {author} {\bibinfo {author} {\bibfnamefont {N.}~\bibnamefont
  {Nagaosa}}, \bibinfo {author} {\bibfnamefont {J.}~\bibnamefont {Sinova}},
  \bibinfo {author} {\bibfnamefont {S.}~\bibnamefont {Onoda}}, \bibinfo
  {author} {\bibfnamefont {A.~H.}\ \bibnamefont {MacDonald}},\ and\ \bibinfo
  {author} {\bibfnamefont {N.~P.}\ \bibnamefont {Ong}},\ }\bibfield  {title}
  {\bibinfo {title} {{Anomalous Hall effect}},\ }\href
  {https://doi.org/10.1103/RevModPhys.82.1539} {\bibfield  {journal} {\bibinfo
  {journal} {Rev. Mod. Phys.}\ }\textbf {\bibinfo {volume} {82}},\ \bibinfo
  {pages} {1539} (\bibinfo {year} {2010})}\BibitemShut {NoStop}%
\bibitem [{\citenamefont {Karplus}\ and\ \citenamefont
  {Luttinger}(1954)}]{Karplus_1954}%
  \BibitemOpen
  \bibfield  {author} {\bibinfo {author} {\bibfnamefont {R.}~\bibnamefont
  {Karplus}}\ and\ \bibinfo {author} {\bibfnamefont {J.~M.}\ \bibnamefont
  {Luttinger}},\ }\bibfield  {title} {\bibinfo {title} {{Hall Effect in
  Ferromagnetics}},\ }\href {https://doi.org/10.1103/PhysRev.95.1154}
  {\bibfield  {journal} {\bibinfo  {journal} {Phys. Rev.}\ }\textbf {\bibinfo
  {volume} {95}},\ \bibinfo {pages} {1154} (\bibinfo {year}
  {1954})}\BibitemShut {NoStop}%
\bibitem [{\citenamefont {Smit}(1958)}]{Smit_1958}%
  \BibitemOpen
  \bibfield  {author} {\bibinfo {author} {\bibfnamefont {J.}~\bibnamefont
  {Smit}},\ }\bibfield  {title} {\bibinfo {title} {{The spontaneous Hall effect
  in ferromagnetics II}},\ }\href
  {https://doi.org/https://doi.org/10.1016/S0031-8914(58)93541-9} {\bibfield
  {journal} {\bibinfo  {journal} {Physica}\ }\textbf {\bibinfo {volume} {24}},\
  \bibinfo {pages} {39} (\bibinfo {year} {1958})}\BibitemShut {NoStop}%
\bibitem [{\citenamefont {Berger}(1970)}]{Berger_1970}%
  \BibitemOpen
  \bibfield  {author} {\bibinfo {author} {\bibfnamefont {L.}~\bibnamefont
  {Berger}},\ }\bibfield  {title} {\bibinfo {title} {{Side-Jump Mechanism for
  the Hall Effect of Ferromagnets}},\ }\href
  {https://doi.org/10.1103/PhysRevB.2.4559} {\bibfield  {journal} {\bibinfo
  {journal} {Phys. Rev. B}\ }\textbf {\bibinfo {volume} {2}},\ \bibinfo {pages}
  {4559} (\bibinfo {year} {1970})}\BibitemShut {NoStop}%
\bibitem [{\citenamefont {Ye}\ \emph {et~al.}(1999)\citenamefont {Ye},
  \citenamefont {Kim}, \citenamefont {Millis}, \citenamefont {Shraiman},
  \citenamefont {Majumdar},\ and\ \citenamefont {Te\ifmmode \check{s}\else
  \v{s}\fi{}anovi\ifmmode~\acute{c}\else \'{c}\fi{}}}]{Ye_AHE_1999}%
  \BibitemOpen
  \bibfield  {author} {\bibinfo {author} {\bibfnamefont {J.}~\bibnamefont
  {Ye}}, \bibinfo {author} {\bibfnamefont {Y.~B.}\ \bibnamefont {Kim}},
  \bibinfo {author} {\bibfnamefont {A.~J.}\ \bibnamefont {Millis}}, \bibinfo
  {author} {\bibfnamefont {B.~I.}\ \bibnamefont {Shraiman}}, \bibinfo {author}
  {\bibfnamefont {P.}~\bibnamefont {Majumdar}},\ and\ \bibinfo {author}
  {\bibfnamefont {Z.}~\bibnamefont {Te\ifmmode \check{s}\else
  \v{s}\fi{}anovi\ifmmode~\acute{c}\else \'{c}\fi{}}},\ }\bibfield  {title}
  {\bibinfo {title} {{Berry Phase Theory of the Anomalous Hall Effect:
  Application to Colossal Magnetoresistance Manganites}},\ }\href
  {https://doi.org/10.1103/PhysRevLett.83.3737} {\bibfield  {journal} {\bibinfo
   {journal} {Phys. Rev. Lett.}\ }\textbf {\bibinfo {volume} {83}},\ \bibinfo
  {pages} {3737} (\bibinfo {year} {1999})}\BibitemShut {NoStop}%
\bibitem [{\citenamefont {Xiao}\ \emph {et~al.}(2010)\citenamefont {Xiao},
  \citenamefont {Chang},\ and\ \citenamefont {Niu}}]{Xiao_RMP_2010}%
  \BibitemOpen
  \bibfield  {author} {\bibinfo {author} {\bibfnamefont {D.}~\bibnamefont
  {Xiao}}, \bibinfo {author} {\bibfnamefont {M.-C.}\ \bibnamefont {Chang}},\
  and\ \bibinfo {author} {\bibfnamefont {Q.}~\bibnamefont {Niu}},\ }\bibfield
  {title} {\bibinfo {title} {{Berry phase effects on electronic properties}},\
  }\href {https://doi.org/10.1103/RevModPhys.82.1959} {\bibfield  {journal}
  {\bibinfo  {journal} {Rev. Mod. Phys.}\ }\textbf {\bibinfo {volume} {82}},\
  \bibinfo {pages} {1959} (\bibinfo {year} {2010})}\BibitemShut {NoStop}%
\bibitem [{\citenamefont {Tomizawa}\ and\ \citenamefont
  {Kontani}(2009)}]{Tomizawa_2009}%
  \BibitemOpen
  \bibfield  {author} {\bibinfo {author} {\bibfnamefont {T.}~\bibnamefont
  {Tomizawa}}\ and\ \bibinfo {author} {\bibfnamefont {H.}~\bibnamefont
  {Kontani}},\ }\bibfield  {title} {\bibinfo {title} {{Anomalous Hall effect in
  the ${t}_{2g}$ orbital kagome lattice due to noncollinearity: Significance of
  the orbital Aharonov-Bohm effect}},\ }\href
  {https://doi.org/10.1103/PhysRevB.80.100401} {\bibfield  {journal} {\bibinfo
  {journal} {Phys. Rev. B}\ }\textbf {\bibinfo {volume} {80}},\ \bibinfo
  {pages} {100401} (\bibinfo {year} {2009})}\BibitemShut {NoStop}%
\bibitem [{\citenamefont {Chen}\ \emph {et~al.}(2014)\citenamefont {Chen},
  \citenamefont {Niu},\ and\ \citenamefont {MacDonald}}]{Chen_2014}%
  \BibitemOpen
  \bibfield  {author} {\bibinfo {author} {\bibfnamefont {H.}~\bibnamefont
  {Chen}}, \bibinfo {author} {\bibfnamefont {Q.}~\bibnamefont {Niu}},\ and\
  \bibinfo {author} {\bibfnamefont {A.~H.}\ \bibnamefont {MacDonald}},\
  }\bibfield  {title} {\bibinfo {title} {{Anomalous Hall Effect Arising from
  Noncollinear Antiferromagnetism}},\ }\href
  {https://doi.org/10.1103/PhysRevLett.112.017205} {\bibfield  {journal}
  {\bibinfo  {journal} {Phys. Rev. Lett.}\ }\textbf {\bibinfo {volume} {112}},\
  \bibinfo {pages} {017205} (\bibinfo {year} {2014})}\BibitemShut {NoStop}%
\bibitem [{\citenamefont {Kübler}\ and\ \citenamefont
  {Felser}(2014)}]{Kubler_2014}%
  \BibitemOpen
  \bibfield  {author} {\bibinfo {author} {\bibfnamefont {J.}~\bibnamefont
  {Kübler}}\ and\ \bibinfo {author} {\bibfnamefont {C.}~\bibnamefont
  {Felser}},\ }\bibfield  {title} {\bibinfo {title} {{Non-collinear
  antiferromagnets and the anomalous Hall effect}},\ }\href
  {https://doi.org/10.1209/0295-5075/108/67001} {\bibfield  {journal} {\bibinfo
   {journal} {{EPL} (Europhysics Letters)}\ }\textbf {\bibinfo {volume}
  {108}},\ \bibinfo {pages} {67001} (\bibinfo {year} {2014})}\BibitemShut
  {NoStop}%
\bibitem [{\citenamefont {Nakatsuji}\ \emph {et~al.}(2015)\citenamefont
  {Nakatsuji}, \citenamefont {Kiyohara},\ and\ \citenamefont
  {Higo}}]{Nakatsuji_2015}%
  \BibitemOpen
  \bibfield  {author} {\bibinfo {author} {\bibfnamefont {S.}~\bibnamefont
  {Nakatsuji}}, \bibinfo {author} {\bibfnamefont {N.}~\bibnamefont
  {Kiyohara}},\ and\ \bibinfo {author} {\bibfnamefont {T.}~\bibnamefont
  {Higo}},\ }\bibfield  {title} {\bibinfo {title} {{Large anomalous Hall effect
  in a non-collinear antiferromagnet at room temperature}},\ }\href
  {https://doi.org/10.1038/nature15723} {\bibfield  {journal} {\bibinfo
  {journal} {Nature}\ }\textbf {\bibinfo {volume} {527}},\ \bibinfo {pages}
  {212} (\bibinfo {year} {2015})}\BibitemShut {NoStop}%
\bibitem [{\citenamefont {Nayak}\ \emph {et~al.}(2016)\citenamefont {Nayak},
  \citenamefont {Fischer}, \citenamefont {Sun}, \citenamefont {Yan},
  \citenamefont {Karel}, \citenamefont {Komarek}, \citenamefont {Shekhar},
  \citenamefont {Kumar}, \citenamefont {Schnelle}, \citenamefont {K{\"u}bler},
  \citenamefont {Felser},\ and\ \citenamefont {Parkin}}]{Nayak_Mn3Ge_2016}%
  \BibitemOpen
  \bibfield  {author} {\bibinfo {author} {\bibfnamefont {A.~K.}\ \bibnamefont
  {Nayak}}, \bibinfo {author} {\bibfnamefont {J.~E.}\ \bibnamefont {Fischer}},
  \bibinfo {author} {\bibfnamefont {Y.}~\bibnamefont {Sun}}, \bibinfo {author}
  {\bibfnamefont {B.}~\bibnamefont {Yan}}, \bibinfo {author} {\bibfnamefont
  {J.}~\bibnamefont {Karel}}, \bibinfo {author} {\bibfnamefont {A.~C.}\
  \bibnamefont {Komarek}}, \bibinfo {author} {\bibfnamefont {C.}~\bibnamefont
  {Shekhar}}, \bibinfo {author} {\bibfnamefont {N.}~\bibnamefont {Kumar}},
  \bibinfo {author} {\bibfnamefont {W.}~\bibnamefont {Schnelle}}, \bibinfo
  {author} {\bibfnamefont {J.}~\bibnamefont {K{\"u}bler}}, \bibinfo {author}
  {\bibfnamefont {C.}~\bibnamefont {Felser}},\ and\ \bibinfo {author}
  {\bibfnamefont {S.~S.~P.}\ \bibnamefont {Parkin}},\ }\bibfield  {title}
  {\bibinfo {title} {{Large anomalous Hall effect driven by a nonvanishing
  Berry curvature in the noncolinear antiferromagnet Mn$_3$Ge}},\ }\href
  {https://doi.org/10.1126/sciadv.1501870} {\bibfield  {journal} {\bibinfo
  {journal} {Science Advances}\ }\textbf {\bibinfo {volume} {2}},\ \bibinfo
  {pages} {e1501870} (\bibinfo {year} {2016})}\BibitemShut {NoStop}%
\bibitem [{\citenamefont {Zhou}\ \emph {et~al.}(2019)\citenamefont {Zhou},
  \citenamefont {Hanke}, \citenamefont {Feng}, \citenamefont {Li},
  \citenamefont {Guo}, \citenamefont {Yao}, \citenamefont {Bl\"ugel},\ and\
  \citenamefont {Mokrousov}}]{Zhou_Mn3XN_2019}%
  \BibitemOpen
  \bibfield  {author} {\bibinfo {author} {\bibfnamefont {X.}~\bibnamefont
  {Zhou}}, \bibinfo {author} {\bibfnamefont {J.-P.}\ \bibnamefont {Hanke}},
  \bibinfo {author} {\bibfnamefont {W.}~\bibnamefont {Feng}}, \bibinfo {author}
  {\bibfnamefont {F.}~\bibnamefont {Li}}, \bibinfo {author} {\bibfnamefont
  {G.-Y.}\ \bibnamefont {Guo}}, \bibinfo {author} {\bibfnamefont
  {Y.}~\bibnamefont {Yao}}, \bibinfo {author} {\bibfnamefont {S.}~\bibnamefont
  {Bl\"ugel}},\ and\ \bibinfo {author} {\bibfnamefont {Y.}~\bibnamefont
  {Mokrousov}},\ }\bibfield  {title} {\bibinfo {title} {{Spin-order dependent
  anomalous Hall effect and magneto-optical effect in the noncollinear
  antiferromagnets ${\mathrm{Mn}}_{3}X\mathrm{N}$ with $X=\mathrm{Ga}$, Zn, Ag,
  or Ni}},\ }\href {https://doi.org/10.1103/PhysRevB.99.104428} {\bibfield
  {journal} {\bibinfo  {journal} {Phys. Rev. B}\ }\textbf {\bibinfo {volume}
  {99}},\ \bibinfo {pages} {104428} (\bibinfo {year} {2019})}\BibitemShut
  {NoStop}%
\bibitem [{\citenamefont {Gurung}\ \emph {et~al.}(2019)\citenamefont {Gurung},
  \citenamefont {Shao}, \citenamefont {Paudel},\ and\ \citenamefont
  {Tsymbal}}]{Gurung_Mn3XN_2019}%
  \BibitemOpen
  \bibfield  {author} {\bibinfo {author} {\bibfnamefont {G.}~\bibnamefont
  {Gurung}}, \bibinfo {author} {\bibfnamefont {D.-F.}\ \bibnamefont {Shao}},
  \bibinfo {author} {\bibfnamefont {T.~R.}\ \bibnamefont {Paudel}},\ and\
  \bibinfo {author} {\bibfnamefont {E.~Y.}\ \bibnamefont {Tsymbal}},\
  }\bibfield  {title} {\bibinfo {title} {{Anomalous Hall conductivity of
  noncollinear magnetic antiperovskites}},\ }\href
  {https://doi.org/10.1103/PhysRevMaterials.3.044409} {\bibfield  {journal}
  {\bibinfo  {journal} {Phys. Rev. Materials}\ }\textbf {\bibinfo {volume}
  {3}},\ \bibinfo {pages} {044409} (\bibinfo {year} {2019})}\BibitemShut
  {NoStop}%
\bibitem [{\citenamefont {Boldrin}\ \emph {et~al.}(2019)\citenamefont
  {Boldrin}, \citenamefont {Samathrakis}, \citenamefont {Zemen}, \citenamefont
  {Mihai}, \citenamefont {Zou}, \citenamefont {Johnson}, \citenamefont {Esser},
  \citenamefont {McComb}, \citenamefont {Petrov}, \citenamefont {Zhang},\ and\
  \citenamefont {Cohen}}]{Boldrin_Mn3XN_2019}%
  \BibitemOpen
  \bibfield  {author} {\bibinfo {author} {\bibfnamefont {D.}~\bibnamefont
  {Boldrin}}, \bibinfo {author} {\bibfnamefont {I.}~\bibnamefont
  {Samathrakis}}, \bibinfo {author} {\bibfnamefont {J.}~\bibnamefont {Zemen}},
  \bibinfo {author} {\bibfnamefont {A.}~\bibnamefont {Mihai}}, \bibinfo
  {author} {\bibfnamefont {B.}~\bibnamefont {Zou}}, \bibinfo {author}
  {\bibfnamefont {F.}~\bibnamefont {Johnson}}, \bibinfo {author} {\bibfnamefont
  {B.~D.}\ \bibnamefont {Esser}}, \bibinfo {author} {\bibfnamefont {D.~W.}\
  \bibnamefont {McComb}}, \bibinfo {author} {\bibfnamefont {P.~K.}\
  \bibnamefont {Petrov}}, \bibinfo {author} {\bibfnamefont {H.}~\bibnamefont
  {Zhang}},\ and\ \bibinfo {author} {\bibfnamefont {L.~F.}\ \bibnamefont
  {Cohen}},\ }\bibfield  {title} {\bibinfo {title} {{Anomalous Hall effect in
  noncollinear antiferromagnetic ${\mathrm{Mn}}_{3}\mathrm{NiN}$ thin films}},\
  }\href {https://doi.org/10.1103/PhysRevMaterials.3.094409} {\bibfield
  {journal} {\bibinfo  {journal} {Phys. Rev. Materials}\ }\textbf {\bibinfo
  {volume} {3}},\ \bibinfo {pages} {094409} (\bibinfo {year}
  {2019})}\BibitemShut {NoStop}%
\bibitem [{\citenamefont {Zhao}\ \emph {et~al.}(2019)\citenamefont {Zhao},
  \citenamefont {Hajiri}, \citenamefont {Chen}, \citenamefont {Miki},
  \citenamefont {Asano},\ and\ \citenamefont {Gegenwart}}]{Zhao_Mn3NiN_2019}%
  \BibitemOpen
  \bibfield  {author} {\bibinfo {author} {\bibfnamefont {K.}~\bibnamefont
  {Zhao}}, \bibinfo {author} {\bibfnamefont {T.}~\bibnamefont {Hajiri}},
  \bibinfo {author} {\bibfnamefont {H.}~\bibnamefont {Chen}}, \bibinfo {author}
  {\bibfnamefont {R.}~\bibnamefont {Miki}}, \bibinfo {author} {\bibfnamefont
  {H.}~\bibnamefont {Asano}},\ and\ \bibinfo {author} {\bibfnamefont
  {P.}~\bibnamefont {Gegenwart}},\ }\bibfield  {title} {\bibinfo {title}
  {{Anomalous Hall effect in the noncollinear antiferromagnetic antiperovskite
  ${\mathrm{Mn}}_{3}{\mathrm{Ni}}_{1\ensuremath{-}x}{\mathrm{Cu}}_{x}\mathrm{N}$}},\
  }\href {https://doi.org/10.1103/PhysRevB.100.045109} {\bibfield  {journal}
  {\bibinfo  {journal} {Phys. Rev. B}\ }\textbf {\bibinfo {volume} {100}},\
  \bibinfo {pages} {045109} (\bibinfo {year} {2019})}\BibitemShut {NoStop}%
\bibitem [{\citenamefont {Liu}\ \emph {et~al.}(2018)\citenamefont {Liu},
  \citenamefont {Chen}, \citenamefont {Wang}, \citenamefont {Liu},
  \citenamefont {Wang}, \citenamefont {Feng}, \citenamefont {Yan},
  \citenamefont {Wang}, \citenamefont {Jiang}, \citenamefont {Coey},\ and\
  \citenamefont {MacDonald}}]{Liu_Mn3Pt_2018}%
  \BibitemOpen
  \bibfield  {author} {\bibinfo {author} {\bibfnamefont {Z.~Q.}\ \bibnamefont
  {Liu}}, \bibinfo {author} {\bibfnamefont {H.}~\bibnamefont {Chen}}, \bibinfo
  {author} {\bibfnamefont {J.~M.}\ \bibnamefont {Wang}}, \bibinfo {author}
  {\bibfnamefont {J.~H.}\ \bibnamefont {Liu}}, \bibinfo {author} {\bibfnamefont
  {K.}~\bibnamefont {Wang}}, \bibinfo {author} {\bibfnamefont {Z.~X.}\
  \bibnamefont {Feng}}, \bibinfo {author} {\bibfnamefont {H.}~\bibnamefont
  {Yan}}, \bibinfo {author} {\bibfnamefont {X.~R.}\ \bibnamefont {Wang}},
  \bibinfo {author} {\bibfnamefont {C.~B.}\ \bibnamefont {Jiang}}, \bibinfo
  {author} {\bibfnamefont {J.~M.~D.}\ \bibnamefont {Coey}},\ and\ \bibinfo
  {author} {\bibfnamefont {A.~H.}\ \bibnamefont {MacDonald}},\ }\bibfield
  {title} {\bibinfo {title} {{Electrical switching of the topological anomalous
  Hall effect in a non-collinear antiferromagnet above room temperature}},\
  }\href {https://doi.org/10.1038/s41928-018-0040-1} {\bibfield  {journal}
  {\bibinfo  {journal} {Nature Electronics}\ }\textbf {\bibinfo {volume} {1}},\
  \bibinfo {pages} {172} (\bibinfo {year} {2018})}\BibitemShut {NoStop}%
\bibitem [{\citenamefont {S{\"u}rgers}\ \emph {et~al.}(2014)\citenamefont
  {S{\"u}rgers}, \citenamefont {Fischer}, \citenamefont {Winkel},\ and\
  \citenamefont {L{\"o}hneysen}}]{Surgers_Mn5Si3_AHE_2014}%
  \BibitemOpen
  \bibfield  {author} {\bibinfo {author} {\bibfnamefont {C.}~\bibnamefont
  {S{\"u}rgers}}, \bibinfo {author} {\bibfnamefont {G.}~\bibnamefont
  {Fischer}}, \bibinfo {author} {\bibfnamefont {P.}~\bibnamefont {Winkel}},\
  and\ \bibinfo {author} {\bibfnamefont {H.~v.}\ \bibnamefont
  {L{\"o}hneysen}},\ }\bibfield  {title} {\bibinfo {title} {{Large topological
  Hall effect in the non-collinear phase of an antiferromagnet}},\ }\href
  {https://doi.org/10.1038/ncomms4400} {\bibfield  {journal} {\bibinfo
  {journal} {Nature Communications}\ }\textbf {\bibinfo {volume} {5}},\
  \bibinfo {pages} {3400} (\bibinfo {year} {2014})}\BibitemShut {NoStop}%
\bibitem [{\citenamefont {{\v S}mejkal}\ \emph {et~al.}(2020)\citenamefont {{\v
  S}mejkal}, \citenamefont {Gonz{\'a}lez-Hern{\'a}ndez}, \citenamefont
  {Jungwirth},\ and\ \citenamefont {Sinova}}]{Smejkal_CHE_2020}%
  \BibitemOpen
  \bibfield  {author} {\bibinfo {author} {\bibfnamefont {L.}~\bibnamefont {{\v
  S}mejkal}}, \bibinfo {author} {\bibfnamefont {R.}~\bibnamefont
  {Gonz{\'a}lez-Hern{\'a}ndez}}, \bibinfo {author} {\bibfnamefont
  {T.}~\bibnamefont {Jungwirth}},\ and\ \bibinfo {author} {\bibfnamefont
  {J.}~\bibnamefont {Sinova}},\ }\bibfield  {title} {\bibinfo {title} {{Crystal
  time-reversal symmetry breaking and spontaneous Hall effect in collinear
  antiferromagnets}},\ }\href {https://doi.org/10.1126/sciadv.aaz8809}
  {\bibfield  {journal} {\bibinfo  {journal} {Science Advances}\ }\textbf
  {\bibinfo {volume} {6}},\ \bibinfo {pages} {eaaz8809} (\bibinfo {year}
  {2020})}\BibitemShut {NoStop}%
\bibitem [{\citenamefont {Li}\ \emph {et~al.}(2019)\citenamefont {Li},
  \citenamefont {MacDonald},\ and\ \citenamefont {Chen}}]{Li_QAHE_2019}%
  \BibitemOpen
  \bibfield  {author} {\bibinfo {author} {\bibfnamefont {X.}~\bibnamefont
  {Li}}, \bibinfo {author} {\bibfnamefont {A.~H.}\ \bibnamefont {MacDonald}},\
  and\ \bibinfo {author} {\bibfnamefont {H.}~\bibnamefont {Chen}},\ }\bibfield
  {title} {\bibinfo {title} {{Quantum Anomalous Hall Effect through Canted
  Antiferromagnetism}},\ }\href@noop {} {\  (\bibinfo {year} {2019})},\ \Eprint
  {https://arxiv.org/abs/arXiv:1902.10650} {arXiv:1902.10650} \BibitemShut
  {NoStop}%
\bibitem [{\citenamefont {Suzuki}\ \emph {et~al.}(2017)\citenamefont {Suzuki},
  \citenamefont {Koretsune}, \citenamefont {Ochi},\ and\ \citenamefont
  {Arita}}]{Suzuki_Cluster_2017}%
  \BibitemOpen
  \bibfield  {author} {\bibinfo {author} {\bibfnamefont {M.-T.}\ \bibnamefont
  {Suzuki}}, \bibinfo {author} {\bibfnamefont {T.}~\bibnamefont {Koretsune}},
  \bibinfo {author} {\bibfnamefont {M.}~\bibnamefont {Ochi}},\ and\ \bibinfo
  {author} {\bibfnamefont {R.}~\bibnamefont {Arita}},\ }\bibfield  {title}
  {\bibinfo {title} {{Cluster multipole theory for anomalous Hall effect in
  antiferromagnets}},\ }\href {https://doi.org/10.1103/PhysRevB.95.094406}
  {\bibfield  {journal} {\bibinfo  {journal} {Phys. Rev. B}\ }\textbf {\bibinfo
  {volume} {95}},\ \bibinfo {pages} {094406} (\bibinfo {year}
  {2017})}\BibitemShut {NoStop}%
\bibitem [{\citenamefont {Hayami}\ and\ \citenamefont
  {Kusunose}(2021)}]{Hayami_AHE_AMD_2021}%
  \BibitemOpen
  \bibfield  {author} {\bibinfo {author} {\bibfnamefont {S.}~\bibnamefont
  {Hayami}}\ and\ \bibinfo {author} {\bibfnamefont {H.}~\bibnamefont
  {Kusunose}},\ }\bibfield  {title} {\bibinfo {title} {{Essential role of the
  anisotropic magnetic dipole in the anomalous Hall effect}},\ }\href
  {https://doi.org/10.1103/PhysRevB.103.L180407} {\bibfield  {journal}
  {\bibinfo  {journal} {Phys. Rev. B}\ }\textbf {\bibinfo {volume} {103}},\
  \bibinfo {pages} {L180407} (\bibinfo {year} {2021})}\BibitemShut {NoStop}%
\bibitem [{\citenamefont {Suzuki}\ \emph {et~al.}(2019)\citenamefont {Suzuki},
  \citenamefont {Nomoto}, \citenamefont {Arita}, \citenamefont {Yanagi},
  \citenamefont {Hayami},\ and\ \citenamefont
  {Kusunose}}]{Suzuki_multipole_2019}%
  \BibitemOpen
  \bibfield  {author} {\bibinfo {author} {\bibfnamefont {M.-T.}\ \bibnamefont
  {Suzuki}}, \bibinfo {author} {\bibfnamefont {T.}~\bibnamefont {Nomoto}},
  \bibinfo {author} {\bibfnamefont {R.}~\bibnamefont {Arita}}, \bibinfo
  {author} {\bibfnamefont {Y.}~\bibnamefont {Yanagi}}, \bibinfo {author}
  {\bibfnamefont {S.}~\bibnamefont {Hayami}},\ and\ \bibinfo {author}
  {\bibfnamefont {H.}~\bibnamefont {Kusunose}},\ }\bibfield  {title} {\bibinfo
  {title} {{Multipole expansion for magnetic structures: A generation scheme
  for a symmetry-adapted orthonormal basis set in the crystallographic point
  group}},\ }\href {https://doi.org/10.1103/PhysRevB.99.174407} {\bibfield
  {journal} {\bibinfo  {journal} {Phys. Rev. B}\ }\textbf {\bibinfo {volume}
  {99}},\ \bibinfo {pages} {174407} (\bibinfo {year} {2019})}\BibitemShut
  {NoStop}%
\bibitem [{\citenamefont {King-Smith}\ and\ \citenamefont
  {Vanderbilt}(1993)}]{king-smith_1993}%
  \BibitemOpen
  \bibfield  {author} {\bibinfo {author} {\bibfnamefont {R.~D.}\ \bibnamefont
  {King-Smith}}\ and\ \bibinfo {author} {\bibfnamefont {D.}~\bibnamefont
  {Vanderbilt}},\ }\bibfield  {title} {\bibinfo {title} {{Theory of
  polarization of crystalline solids}},\ }\href
  {https://doi.org/10.1103/PhysRevB.47.1651} {\bibfield  {journal} {\bibinfo
  {journal} {Phys. Rev. B}\ }\textbf {\bibinfo {volume} {47}},\ \bibinfo
  {pages} {1651} (\bibinfo {year} {1993})}\BibitemShut {NoStop}%
\bibitem [{\citenamefont {Resta}(1994)}]{Resta_1994}%
  \BibitemOpen
  \bibfield  {author} {\bibinfo {author} {\bibfnamefont {R.}~\bibnamefont
  {Resta}},\ }\bibfield  {title} {\bibinfo {title} {{Macroscopic polarization
  in crystalline dielectrics: the geometric phase approach}},\ }\href
  {https://doi.org/10.1103/RevModPhys.66.899} {\bibfield  {journal} {\bibinfo
  {journal} {Rev. Mod. Phys.}\ }\textbf {\bibinfo {volume} {66}},\ \bibinfo
  {pages} {899} (\bibinfo {year} {1994})}\BibitemShut {NoStop}%
\bibitem [{\citenamefont {Xiao}\ \emph {et~al.}(2005)\citenamefont {Xiao},
  \citenamefont {Shi},\ and\ \citenamefont {Niu}}]{xiao_2005}%
  \BibitemOpen
  \bibfield  {author} {\bibinfo {author} {\bibfnamefont {D.}~\bibnamefont
  {Xiao}}, \bibinfo {author} {\bibfnamefont {J.}~\bibnamefont {Shi}},\ and\
  \bibinfo {author} {\bibfnamefont {Q.}~\bibnamefont {Niu}},\ }\bibfield
  {title} {\bibinfo {title} {{Berry Phase Correction to Electron Density of
  States in Solids}},\ }\href {https://doi.org/10.1103/PhysRevLett.95.137204}
  {\bibfield  {journal} {\bibinfo  {journal} {Phys. Rev. Lett.}\ }\textbf
  {\bibinfo {volume} {95}},\ \bibinfo {pages} {137204} (\bibinfo {year}
  {2005})}\BibitemShut {NoStop}%
\bibitem [{\citenamefont {Thonhauser}\ \emph {et~al.}(2005)\citenamefont
  {Thonhauser}, \citenamefont {Ceresoli}, \citenamefont {Vanderbilt},\ and\
  \citenamefont {Resta}}]{thonhauser_2005}%
  \BibitemOpen
  \bibfield  {author} {\bibinfo {author} {\bibfnamefont {T.}~\bibnamefont
  {Thonhauser}}, \bibinfo {author} {\bibfnamefont {D.}~\bibnamefont
  {Ceresoli}}, \bibinfo {author} {\bibfnamefont {D.}~\bibnamefont
  {Vanderbilt}},\ and\ \bibinfo {author} {\bibfnamefont {R.}~\bibnamefont
  {Resta}},\ }\bibfield  {title} {\bibinfo {title} {{Orbital Magnetization in
  Periodic Insulators}},\ }\href
  {https://doi.org/10.1103/PhysRevLett.95.137205} {\bibfield  {journal}
  {\bibinfo  {journal} {Phys. Rev. Lett.}\ }\textbf {\bibinfo {volume} {95}},\
  \bibinfo {pages} {137205} (\bibinfo {year} {2005})}\BibitemShut {NoStop}%
\bibitem [{\citenamefont {Shi}\ \emph {et~al.}(2007)\citenamefont {Shi},
  \citenamefont {Vignale}, \citenamefont {Xiao},\ and\ \citenamefont
  {Niu}}]{shi_2007}%
  \BibitemOpen
  \bibfield  {author} {\bibinfo {author} {\bibfnamefont {J.}~\bibnamefont
  {Shi}}, \bibinfo {author} {\bibfnamefont {G.}~\bibnamefont {Vignale}},
  \bibinfo {author} {\bibfnamefont {D.}~\bibnamefont {Xiao}},\ and\ \bibinfo
  {author} {\bibfnamefont {Q.}~\bibnamefont {Niu}},\ }\bibfield  {title}
  {\bibinfo {title} {{Quantum Theory of Orbital Magnetization and Its
  Generalization to Interacting Systems}},\ }\href
  {https://doi.org/10.1103/PhysRevLett.99.197202} {\bibfield  {journal}
  {\bibinfo  {journal} {Phys. Rev. Lett.}\ }\textbf {\bibinfo {volume} {99}},\
  \bibinfo {pages} {197202} (\bibinfo {year} {2007})}\BibitemShut {NoStop}%
\bibitem [{\citenamefont {Spaldin}\ \emph {et~al.}(2008)\citenamefont
  {Spaldin}, \citenamefont {Fiebig},\ and\ \citenamefont
  {Mostovoy}}]{Spaldin_2008}%
  \BibitemOpen
  \bibfield  {author} {\bibinfo {author} {\bibfnamefont {N.~A.}\ \bibnamefont
  {Spaldin}}, \bibinfo {author} {\bibfnamefont {M.}~\bibnamefont {Fiebig}},\
  and\ \bibinfo {author} {\bibfnamefont {M.}~\bibnamefont {Mostovoy}},\
  }\bibfield  {title} {\bibinfo {title} {{The toroidal moment in
  condensed-matter physics and its relation to the magnetoelectric effect}},\
  }\href {https://doi.org/10.1088/0953-8984/20/43/434203} {\bibfield  {journal}
  {\bibinfo  {journal} {Journal of Physics: Condensed Matter}\ }\textbf
  {\bibinfo {volume} {20}},\ \bibinfo {pages} {434203} (\bibinfo {year}
  {2008})}\BibitemShut {NoStop}%
\bibitem [{\citenamefont {Gao}\ \emph {et~al.}(2018)\citenamefont {Gao},
  \citenamefont {Vanderbilt},\ and\ \citenamefont {Xiao}}]{gao_2018}%
  \BibitemOpen
  \bibfield  {author} {\bibinfo {author} {\bibfnamefont {Y.}~\bibnamefont
  {Gao}}, \bibinfo {author} {\bibfnamefont {D.}~\bibnamefont {Vanderbilt}},\
  and\ \bibinfo {author} {\bibfnamefont {D.}~\bibnamefont {Xiao}},\ }\bibfield
  {title} {\bibinfo {title} {{Microscopic theory of spin toroidization in
  periodic crystals}},\ }\href {https://doi.org/10.1103/PhysRevB.97.134423}
  {\bibfield  {journal} {\bibinfo  {journal} {Phys. Rev. B}\ }\textbf {\bibinfo
  {volume} {97}},\ \bibinfo {pages} {134423} (\bibinfo {year}
  {2018})}\BibitemShut {NoStop}%
\bibitem [{\citenamefont {Lipkin}(1964)}]{lipkin_1964}%
  \BibitemOpen
  \bibfield  {author} {\bibinfo {author} {\bibfnamefont {D.~M.}\ \bibnamefont
  {Lipkin}},\ }\bibfield  {title} {\bibinfo {title} {{Existence of a New
  Conservation Law in Electromagnetic Theory}},\ }\href
  {https://doi.org/10.1063/1.1704165} {\bibfield  {journal} {\bibinfo
  {journal} {Journal of Mathematical Physics}\ }\textbf {\bibinfo {volume}
  {5}},\ \bibinfo {pages} {696} (\bibinfo {year} {1964})},\ \Eprint
  {https://arxiv.org/abs/https://doi.org/10.1063/1.1704165}
  {https://doi.org/10.1063/1.1704165} \BibitemShut {NoStop}%
\bibitem [{\citenamefont {Tang}\ and\ \citenamefont {Cohen}(2010)}]{tang_2010}%
  \BibitemOpen
  \bibfield  {author} {\bibinfo {author} {\bibfnamefont {Y.}~\bibnamefont
  {Tang}}\ and\ \bibinfo {author} {\bibfnamefont {A.~E.}\ \bibnamefont
  {Cohen}},\ }\bibfield  {title} {\bibinfo {title} {{Optical Chirality and Its
  Interaction with Matter}},\ }\href
  {https://doi.org/10.1103/PhysRevLett.104.163901} {\bibfield  {journal}
  {\bibinfo  {journal} {Phys. Rev. Lett.}\ }\textbf {\bibinfo {volume} {104}},\
  \bibinfo {pages} {163901} (\bibinfo {year} {2010})}\BibitemShut {NoStop}%
\bibitem [{\citenamefont {Bliokh}\ and\ \citenamefont
  {Nori}(2011)}]{bliokh_2011}%
  \BibitemOpen
  \bibfield  {author} {\bibinfo {author} {\bibfnamefont {K.~Y.}\ \bibnamefont
  {Bliokh}}\ and\ \bibinfo {author} {\bibfnamefont {F.}~\bibnamefont {Nori}},\
  }\bibfield  {title} {\bibinfo {title} {{Characterizing optical chirality}},\
  }\href {https://doi.org/10.1103/PhysRevA.83.021803} {\bibfield  {journal}
  {\bibinfo  {journal} {Phys. Rev. A}\ }\textbf {\bibinfo {volume} {83}},\
  \bibinfo {pages} {021803} (\bibinfo {year} {2011})}\BibitemShut {NoStop}%
\bibitem [{\citenamefont {Coles}\ and\ \citenamefont
  {Andrews}(2012)}]{coles_2012}%
  \BibitemOpen
  \bibfield  {author} {\bibinfo {author} {\bibfnamefont {M.~M.}\ \bibnamefont
  {Coles}}\ and\ \bibinfo {author} {\bibfnamefont {D.~L.}\ \bibnamefont
  {Andrews}},\ }\bibfield  {title} {\bibinfo {title} {{Chirality and angular
  momentum in optical radiation}},\ }\href
  {https://doi.org/10.1103/PhysRevA.85.063810} {\bibfield  {journal} {\bibinfo
  {journal} {Phys. Rev. A}\ }\textbf {\bibinfo {volume} {85}},\ \bibinfo
  {pages} {063810} (\bibinfo {year} {2012})}\BibitemShut {NoStop}%
\bibitem [{\citenamefont {Wills}\ \emph {et~al.}(2002)\citenamefont {Wills},
  \citenamefont {Ballou},\ and\ \citenamefont {Lacroix}}]{wills_2002}%
  \BibitemOpen
  \bibfield  {author} {\bibinfo {author} {\bibfnamefont {A.~S.}\ \bibnamefont
  {Wills}}, \bibinfo {author} {\bibfnamefont {R.}~\bibnamefont {Ballou}},\ and\
  \bibinfo {author} {\bibfnamefont {C.}~\bibnamefont {Lacroix}},\ }\bibfield
  {title} {\bibinfo {title} {{Model of localized highly frustrated
  ferromagnetism: The kagom\'e spin ice}},\ }\href
  {https://doi.org/10.1103/PhysRevB.66.144407} {\bibfield  {journal} {\bibinfo
  {journal} {Phys. Rev. B}\ }\textbf {\bibinfo {volume} {66}},\ \bibinfo
  {pages} {144407} (\bibinfo {year} {2002})}\BibitemShut {NoStop}%
\bibitem [{\citenamefont {M\"oller}\ and\ \citenamefont
  {Moessner}(2009)}]{moller_2009}%
  \BibitemOpen
  \bibfield  {author} {\bibinfo {author} {\bibfnamefont {G.}~\bibnamefont
  {M\"oller}}\ and\ \bibinfo {author} {\bibfnamefont {R.}~\bibnamefont
  {Moessner}},\ }\bibfield  {title} {\bibinfo {title} {{Magnetic multipole
  analysis of kagome and artificial spin-ice dipolar arrays}},\ }\href
  {https://doi.org/10.1103/PhysRevB.80.140409} {\bibfield  {journal} {\bibinfo
  {journal} {Phys. Rev. B}\ }\textbf {\bibinfo {volume} {80}},\ \bibinfo
  {pages} {140409} (\bibinfo {year} {2009})}\BibitemShut {NoStop}%
\bibitem [{\citenamefont {Chern}\ \emph {et~al.}(2011)\citenamefont {Chern},
  \citenamefont {Mellado},\ and\ \citenamefont {Tchernyshyov}}]{chern_2011}%
  \BibitemOpen
  \bibfield  {author} {\bibinfo {author} {\bibfnamefont {G.-W.}\ \bibnamefont
  {Chern}}, \bibinfo {author} {\bibfnamefont {P.}~\bibnamefont {Mellado}},\
  and\ \bibinfo {author} {\bibfnamefont {O.}~\bibnamefont {Tchernyshyov}},\
  }\bibfield  {title} {\bibinfo {title} {{Two-Stage Ordering of Spins in
  Dipolar Spin Ice on the Kagome Lattice}},\ }\href
  {https://doi.org/10.1103/PhysRevLett.106.207202} {\bibfield  {journal}
  {\bibinfo  {journal} {Phys. Rev. Lett.}\ }\textbf {\bibinfo {volume} {106}},\
  \bibinfo {pages} {207202} (\bibinfo {year} {2011})}\BibitemShut {NoStop}%
\bibitem [{\citenamefont {Qi}\ \emph {et~al.}(2008)\citenamefont {Qi},
  \citenamefont {Brintlinger},\ and\ \citenamefont {Cumings}}]{qi_2008}%
  \BibitemOpen
  \bibfield  {author} {\bibinfo {author} {\bibfnamefont {Y.}~\bibnamefont
  {Qi}}, \bibinfo {author} {\bibfnamefont {T.}~\bibnamefont {Brintlinger}},\
  and\ \bibinfo {author} {\bibfnamefont {J.}~\bibnamefont {Cumings}},\
  }\bibfield  {title} {\bibinfo {title} {{Direct observation of the ice rule in
  an artificial kagome spin ice}},\ }\href
  {https://doi.org/10.1103/PhysRevB.77.094418} {\bibfield  {journal} {\bibinfo
  {journal} {Phys. Rev. B}\ }\textbf {\bibinfo {volume} {77}},\ \bibinfo
  {pages} {094418} (\bibinfo {year} {2008})}\BibitemShut {NoStop}%
\bibitem [{\citenamefont {Ladak}\ \emph {et~al.}(2010)\citenamefont {Ladak},
  \citenamefont {Read}, \citenamefont {Perkins}, \citenamefont {Cohen},\ and\
  \citenamefont {Branford}}]{ladak_2010}%
  \BibitemOpen
  \bibfield  {author} {\bibinfo {author} {\bibfnamefont {S.}~\bibnamefont
  {Ladak}}, \bibinfo {author} {\bibfnamefont {D.~E.}\ \bibnamefont {Read}},
  \bibinfo {author} {\bibfnamefont {G.~K.}\ \bibnamefont {Perkins}}, \bibinfo
  {author} {\bibfnamefont {L.~F.}\ \bibnamefont {Cohen}},\ and\ \bibinfo
  {author} {\bibfnamefont {W.~R.}\ \bibnamefont {Branford}},\ }\bibfield
  {title} {\bibinfo {title} {{Direct observation of magnetic monopole defects
  in an artificial spin-ice system}},\ }\href
  {https://doi.org/10.1038/nphys1628} {\bibfield  {journal} {\bibinfo
  {journal} {Nature Physics}\ }\textbf {\bibinfo {volume} {6}},\ \bibinfo
  {pages} {359} (\bibinfo {year} {2010})}\BibitemShut {NoStop}%
\bibitem [{\citenamefont {Dun}\ \emph {et~al.}(2016)\citenamefont {Dun},
  \citenamefont {Trinh}, \citenamefont {Li}, \citenamefont {Lee}, \citenamefont
  {Chen}, \citenamefont {Baumbach}, \citenamefont {Hu}, \citenamefont {Wang},
  \citenamefont {Choi}, \citenamefont {Shastry}, \citenamefont {Ramirez},\ and\
  \citenamefont {Zhou}}]{dun_2016}%
  \BibitemOpen
  \bibfield  {author} {\bibinfo {author} {\bibfnamefont {Z.~L.}\ \bibnamefont
  {Dun}}, \bibinfo {author} {\bibfnamefont {J.}~\bibnamefont {Trinh}}, \bibinfo
  {author} {\bibfnamefont {K.}~\bibnamefont {Li}}, \bibinfo {author}
  {\bibfnamefont {M.}~\bibnamefont {Lee}}, \bibinfo {author} {\bibfnamefont
  {K.~W.}\ \bibnamefont {Chen}}, \bibinfo {author} {\bibfnamefont
  {R.}~\bibnamefont {Baumbach}}, \bibinfo {author} {\bibfnamefont {Y.~F.}\
  \bibnamefont {Hu}}, \bibinfo {author} {\bibfnamefont {Y.~X.}\ \bibnamefont
  {Wang}}, \bibinfo {author} {\bibfnamefont {E.~S.}\ \bibnamefont {Choi}},
  \bibinfo {author} {\bibfnamefont {B.~S.}\ \bibnamefont {Shastry}}, \bibinfo
  {author} {\bibfnamefont {A.~P.}\ \bibnamefont {Ramirez}},\ and\ \bibinfo
  {author} {\bibfnamefont {H.~D.}\ \bibnamefont {Zhou}},\ }\bibfield  {title}
  {\bibinfo {title} {{Magnetic Ground States of the Rare-Earth Tripod Kagome
  Lattice
  ${\mathrm{Mg}}_{2}{\mathrm{RE}}_{3}{\mathrm{Sb}}_{3}{\mathrm{O}}_{14}$
  ($\mathrm{RE}=\mathrm{Gd},\mathrm{Dy},\mathrm{Er}$)}},\ }\href
  {https://doi.org/10.1103/PhysRevLett.116.157201} {\bibfield  {journal}
  {\bibinfo  {journal} {Phys. Rev. Lett.}\ }\textbf {\bibinfo {volume} {116}},\
  \bibinfo {pages} {157201} (\bibinfo {year} {2016})}\BibitemShut {NoStop}%
\bibitem [{\citenamefont {Matsuhira}\ \emph {et~al.}(2002)\citenamefont
  {Matsuhira}, \citenamefont {Hiroi}, \citenamefont {Tayama}, \citenamefont
  {Takagi},\ and\ \citenamefont {Sakakibara}}]{Matsuhira_2002}%
  \BibitemOpen
  \bibfield  {author} {\bibinfo {author} {\bibfnamefont {K.}~\bibnamefont
  {Matsuhira}}, \bibinfo {author} {\bibfnamefont {Z.}~\bibnamefont {Hiroi}},
  \bibinfo {author} {\bibfnamefont {T.}~\bibnamefont {Tayama}}, \bibinfo
  {author} {\bibfnamefont {S.}~\bibnamefont {Takagi}},\ and\ \bibinfo {author}
  {\bibfnamefont {T.}~\bibnamefont {Sakakibara}},\ }\bibfield  {title}
  {\bibinfo {title} {{A new macroscopically degenerate ground state in the spin
  ice compound Dy$_2$Ti$_2$O$_7$ under a magnetic field}},\ }\href
  {https://doi.org/10.1088/0953-8984/14/29/101} {\bibfield  {journal} {\bibinfo
   {journal} {Journal of Physics: Condensed Matter}\ }\textbf {\bibinfo
  {volume} {14}},\ \bibinfo {pages} {L559} (\bibinfo {year}
  {2002})}\BibitemShut {NoStop}%
\bibitem [{\citenamefont {Tabata}\ \emph {et~al.}(2006)\citenamefont {Tabata},
  \citenamefont {Kadowaki}, \citenamefont {Matsuhira}, \citenamefont {Hiroi},
  \citenamefont {Aso}, \citenamefont {Ressouche},\ and\ \citenamefont
  {F\aa{}k}}]{tabata_2006}%
  \BibitemOpen
  \bibfield  {author} {\bibinfo {author} {\bibfnamefont {Y.}~\bibnamefont
  {Tabata}}, \bibinfo {author} {\bibfnamefont {H.}~\bibnamefont {Kadowaki}},
  \bibinfo {author} {\bibfnamefont {K.}~\bibnamefont {Matsuhira}}, \bibinfo
  {author} {\bibfnamefont {Z.}~\bibnamefont {Hiroi}}, \bibinfo {author}
  {\bibfnamefont {N.}~\bibnamefont {Aso}}, \bibinfo {author} {\bibfnamefont
  {E.}~\bibnamefont {Ressouche}},\ and\ \bibinfo {author} {\bibfnamefont
  {B.}~\bibnamefont {F\aa{}k}},\ }\bibfield  {title} {\bibinfo {title}
  {{Kagom\'e Ice State in the Dipolar Spin Ice
  ${\mathrm{Dy}}_{2}{\mathrm{Ti}}_{2}{\mathrm{O}}_{7}$}},\ }\href
  {https://doi.org/10.1103/PhysRevLett.97.257205} {\bibfield  {journal}
  {\bibinfo  {journal} {Phys. Rev. Lett.}\ }\textbf {\bibinfo {volume} {97}},\
  \bibinfo {pages} {257205} (\bibinfo {year} {2006})}\BibitemShut {NoStop}%
\bibitem [{\citenamefont {Takagi}\ and\ \citenamefont
  {Mekata}(1993)}]{takagi_1993}%
  \BibitemOpen
  \bibfield  {author} {\bibinfo {author} {\bibfnamefont {T.}~\bibnamefont
  {Takagi}}\ and\ \bibinfo {author} {\bibfnamefont {M.}~\bibnamefont
  {Mekata}},\ }\bibfield  {title} {\bibinfo {title} {{Magnetic Ordering of
  Ising Spins on Kagomé Lattice with the 1st and the 2nd Neighbor
  Interactions}},\ }\href {https://doi.org/10.1143/JPSJ.62.3943} {\bibfield
  {journal} {\bibinfo  {journal} {Journal of the Physical Society of Japan}\
  }\textbf {\bibinfo {volume} {62}},\ \bibinfo {pages} {3943} (\bibinfo {year}
  {1993})},\ \Eprint
  {https://arxiv.org/abs/https://doi.org/10.1143/JPSJ.62.3943}
  {https://doi.org/10.1143/JPSJ.62.3943} \BibitemShut {NoStop}%
\bibitem [{\citenamefont {Chern}\ and\ \citenamefont
  {Tchernyshyov}(2012)}]{chern_2012}%
  \BibitemOpen
  \bibfield  {author} {\bibinfo {author} {\bibfnamefont {G.-W.}\ \bibnamefont
  {Chern}}\ and\ \bibinfo {author} {\bibfnamefont {O.}~\bibnamefont
  {Tchernyshyov}},\ }\bibfield  {title} {\bibinfo {title} {{Magnetic charge and
  ordering in kagome spin ice}},\ }\href
  {https://doi.org/10.1098/rsta.2011.0388} {\bibfield  {journal} {\bibinfo
  {journal} {Philosophical Transactions of the Royal Society A: Mathematical,
  Physical and Engineering Sciences}\ }\textbf {\bibinfo {volume} {370}},\
  \bibinfo {pages} {5718} (\bibinfo {year} {2012})}\BibitemShut {NoStop}%
\bibitem [{\citenamefont {Wolf}\ and\ \citenamefont
  {Schotte}(1988)}]{Wolf_1988}%
  \BibitemOpen
  \bibfield  {author} {\bibinfo {author} {\bibfnamefont {M.}~\bibnamefont
  {Wolf}}\ and\ \bibinfo {author} {\bibfnamefont {K.~D.}\ \bibnamefont
  {Schotte}},\ }\bibfield  {title} {\bibinfo {title} {{Ising model with
  competing next-nearest-neighbour interactions on the Kagome lattice}},\
  }\href {https://doi.org/10.1088/0305-4470/21/9/032} {\bibfield  {journal}
  {\bibinfo  {journal} {Journal of Physics A: Mathematical and General}\
  }\textbf {\bibinfo {volume} {21}},\ \bibinfo {pages} {2195} (\bibinfo {year}
  {1988})}\BibitemShut {NoStop}%
\bibitem [{\citenamefont {Zhao}\ \emph {et~al.}(2020)\citenamefont {Zhao},
  \citenamefont {Deng}, \citenamefont {Chen}, \citenamefont {Ross},
  \citenamefont {Petříček}, \citenamefont {Günther}, \citenamefont
  {Russina}, \citenamefont {Hutanu},\ and\ \citenamefont
  {Gegenwart}}]{zhao_2020}%
  \BibitemOpen
  \bibfield  {author} {\bibinfo {author} {\bibfnamefont {K.}~\bibnamefont
  {Zhao}}, \bibinfo {author} {\bibfnamefont {H.}~\bibnamefont {Deng}}, \bibinfo
  {author} {\bibfnamefont {H.}~\bibnamefont {Chen}}, \bibinfo {author}
  {\bibfnamefont {K.~A.}\ \bibnamefont {Ross}}, \bibinfo {author}
  {\bibfnamefont {V.}~\bibnamefont {Petříček}}, \bibinfo {author}
  {\bibfnamefont {G.}~\bibnamefont {Günther}}, \bibinfo {author}
  {\bibfnamefont {M.}~\bibnamefont {Russina}}, \bibinfo {author} {\bibfnamefont
  {V.}~\bibnamefont {Hutanu}},\ and\ \bibinfo {author} {\bibfnamefont
  {P.}~\bibnamefont {Gegenwart}},\ }\bibfield  {title} {\bibinfo {title}
  {{Realization of the kagome spin ice state in a frustrated intermetallic
  compound}},\ }\href {https://doi.org/10.1126/science.aaw1666} {\bibfield
  {journal} {\bibinfo  {journal} {Science}\ }\textbf {\bibinfo {volume}
  {367}},\ \bibinfo {pages} {1218} (\bibinfo {year} {2020})}\BibitemShut
  {NoStop}%
\bibitem [{\citenamefont {Lovesey}(1986)}]{lovesey_book}%
  \BibitemOpen
  \bibfield  {author} {\bibinfo {author} {\bibfnamefont {S.}~\bibnamefont
  {Lovesey}},\ }\href
  {https://global.oup.com/academic/product/theory-of-neutron-scattering-from-condensed-matter-9780198520290}
  {\emph {\bibinfo {title} {{Theory of Neutron Scattering from Condensed
  Matter, Volume II: Polarization Effects and Magnetic Scattering}}}}\
  (\bibinfo  {publisher} {Clarendon Press},\ \bibinfo {year}
  {1986})\BibitemShut {NoStop}%
\bibitem [{\citenamefont {Bychkov}\ and\ \citenamefont
  {Rashba}(1984)}]{Bychkov_1984}%
  \BibitemOpen
  \bibfield  {author} {\bibinfo {author} {\bibfnamefont {Y.~A.}\ \bibnamefont
  {Bychkov}}\ and\ \bibinfo {author} {\bibfnamefont {E.~I.}\ \bibnamefont
  {Rashba}},\ }\bibfield  {title} {\bibinfo {title} {{Oscillatory effects and
  the magnetic susceptibility of carriers in inversion layers}},\ }\href
  {https://doi.org/10.1088/0022-3719/17/33/015} {\bibfield  {journal} {\bibinfo
   {journal} {Journal of Physics C: Solid State Physics}\ }\textbf {\bibinfo
  {volume} {17}},\ \bibinfo {pages} {6039} (\bibinfo {year}
  {1984})}\BibitemShut {NoStop}%
\bibitem [{\citenamefont {Chen}\ \emph {et~al.}(2020)\citenamefont {Chen},
  \citenamefont {Wang}, \citenamefont {Xiao}, \citenamefont {Guo},
  \citenamefont {Niu},\ and\ \citenamefont {MacDonald}}]{Chen_2020}%
  \BibitemOpen
  \bibfield  {author} {\bibinfo {author} {\bibfnamefont {H.}~\bibnamefont
  {Chen}}, \bibinfo {author} {\bibfnamefont {T.-C.}\ \bibnamefont {Wang}},
  \bibinfo {author} {\bibfnamefont {D.}~\bibnamefont {Xiao}}, \bibinfo {author}
  {\bibfnamefont {G.-Y.}\ \bibnamefont {Guo}}, \bibinfo {author} {\bibfnamefont
  {Q.}~\bibnamefont {Niu}},\ and\ \bibinfo {author} {\bibfnamefont {A.~H.}\
  \bibnamefont {MacDonald}},\ }\bibfield  {title} {\bibinfo {title}
  {{Manipulating anomalous Hall antiferromagnets with magnetic fields}},\
  }\href {https://doi.org/10.1103/PhysRevB.101.104418} {\bibfield  {journal}
  {\bibinfo  {journal} {Phys. Rev. B}\ }\textbf {\bibinfo {volume} {101}},\
  \bibinfo {pages} {104418} (\bibinfo {year} {2020})}\BibitemShut {NoStop}%
\bibitem [{\citenamefont {Birss}(1964)}]{birss1964symmetry}%
  \BibitemOpen
  \bibfield  {author} {\bibinfo {author} {\bibfnamefont {R.}~\bibnamefont
  {Birss}},\ }\href@noop {} {\emph {\bibinfo {title} {{Symmetry and
  magnetism}}}},\ Selected topics in solid state physics\ (\bibinfo
  {publisher} {North-Holland Pub. Co.},\ \bibinfo {year} {1964})\BibitemShut
  {NoStop}%
\bibitem [{\citenamefont {Tse}\ \emph {et~al.}(2011)\citenamefont {Tse},
  \citenamefont {Qiao}, \citenamefont {Yao}, \citenamefont {MacDonald},\ and\
  \citenamefont {Niu}}]{tse_2011}%
  \BibitemOpen
  \bibfield  {author} {\bibinfo {author} {\bibfnamefont {W.-K.}\ \bibnamefont
  {Tse}}, \bibinfo {author} {\bibfnamefont {Z.}~\bibnamefont {Qiao}}, \bibinfo
  {author} {\bibfnamefont {Y.}~\bibnamefont {Yao}}, \bibinfo {author}
  {\bibfnamefont {A.~H.}\ \bibnamefont {MacDonald}},\ and\ \bibinfo {author}
  {\bibfnamefont {Q.}~\bibnamefont {Niu}},\ }\bibfield  {title} {\bibinfo
  {title} {{Quantum anomalous Hall effect in single-layer and bilayer
  graphene}},\ }\href {https://doi.org/10.1103/PhysRevB.83.155447} {\bibfield
  {journal} {\bibinfo  {journal} {Phys. Rev. B}\ }\textbf {\bibinfo {volume}
  {83}},\ \bibinfo {pages} {155447} (\bibinfo {year} {2011})}\BibitemShut
  {NoStop}%
\bibitem [{\citenamefont {Ishizuka}\ and\ \citenamefont
  {Motome}(2013)}]{ishizuka_2013}%
  \BibitemOpen
  \bibfield  {author} {\bibinfo {author} {\bibfnamefont {H.}~\bibnamefont
  {Ishizuka}}\ and\ \bibinfo {author} {\bibfnamefont {Y.}~\bibnamefont
  {Motome}},\ }\bibfield  {title} {\bibinfo {title} {{Quantum anomalous Hall
  effect in kagome ice}},\ }\href {https://doi.org/10.1103/PhysRevB.87.081105}
  {\bibfield  {journal} {\bibinfo  {journal} {Phys. Rev. B}\ }\textbf {\bibinfo
  {volume} {87}},\ \bibinfo {pages} {081105} (\bibinfo {year}
  {2013})}\BibitemShut {NoStop}%
\bibitem [{\citenamefont {Chern}\ \emph {et~al.}(2014)\citenamefont {Chern},
  \citenamefont {Rahmani}, \citenamefont {Martin},\ and\ \citenamefont
  {Batista}}]{chern_2014}%
  \BibitemOpen
  \bibfield  {author} {\bibinfo {author} {\bibfnamefont {G.-W.}\ \bibnamefont
  {Chern}}, \bibinfo {author} {\bibfnamefont {A.}~\bibnamefont {Rahmani}},
  \bibinfo {author} {\bibfnamefont {I.}~\bibnamefont {Martin}},\ and\ \bibinfo
  {author} {\bibfnamefont {C.~D.}\ \bibnamefont {Batista}},\ }\bibfield
  {title} {\bibinfo {title} {{Quantum Hall ice}},\ }\href
  {https://doi.org/10.1103/PhysRevB.90.241102} {\bibfield  {journal} {\bibinfo
  {journal} {Phys. Rev. B}\ }\textbf {\bibinfo {volume} {90}},\ \bibinfo
  {pages} {241102} (\bibinfo {year} {2014})}\BibitemShut {NoStop}%
\bibitem [{\citenamefont {Machida}\ \emph {et~al.}(2010)\citenamefont
  {Machida}, \citenamefont {Nakatsuji}, \citenamefont {Onoda}, \citenamefont
  {Tayama},\ and\ \citenamefont {Sakakibara}}]{Machida_AHE_no_order_2010}%
  \BibitemOpen
  \bibfield  {author} {\bibinfo {author} {\bibfnamefont {Y.}~\bibnamefont
  {Machida}}, \bibinfo {author} {\bibfnamefont {S.}~\bibnamefont {Nakatsuji}},
  \bibinfo {author} {\bibfnamefont {S.}~\bibnamefont {Onoda}}, \bibinfo
  {author} {\bibfnamefont {T.}~\bibnamefont {Tayama}},\ and\ \bibinfo {author}
  {\bibfnamefont {T.}~\bibnamefont {Sakakibara}},\ }\bibfield  {title}
  {\bibinfo {title} {{Time-reversal symmetry breaking and spontaneous Hall
  effect without magnetic dipole order}},\ }\href
  {https://doi.org/10.1038/nature08680} {\bibfield  {journal} {\bibinfo
  {journal} {Nature}\ }\textbf {\bibinfo {volume} {463}},\ \bibinfo {pages}
  {210} (\bibinfo {year} {2010})}\BibitemShut {NoStop}%
\bibitem [{\citenamefont {van~der Laan}(2021)}]{vanderLaan_2021}%
  \BibitemOpen
  \bibfield  {author} {\bibinfo {author} {\bibfnamefont {G.}~\bibnamefont
  {van~der Laan}},\ }\bibfield  {title} {\bibinfo {title} {{Determination of
  spin chirality using x-ray magnetic circular dichroism}},\ }\href
  {https://doi.org/10.1103/PhysRevB.104.094414} {\bibfield  {journal} {\bibinfo
   {journal} {Phys. Rev. B}\ }\textbf {\bibinfo {volume} {104}},\ \bibinfo
  {pages} {094414} (\bibinfo {year} {2021})}\BibitemShut {NoStop}%
\bibitem [{\citenamefont {Ferreira}\ \emph {et~al.}(2014)\citenamefont
  {Ferreira}, \citenamefont {Rappoport}, \citenamefont {Cazalilla},\ and\
  \citenamefont {Castro~Neto}}]{ferreira_2014}%
  \BibitemOpen
  \bibfield  {author} {\bibinfo {author} {\bibfnamefont {A.}~\bibnamefont
  {Ferreira}}, \bibinfo {author} {\bibfnamefont {T.~G.}\ \bibnamefont
  {Rappoport}}, \bibinfo {author} {\bibfnamefont {M.~A.}\ \bibnamefont
  {Cazalilla}},\ and\ \bibinfo {author} {\bibfnamefont {A.~H.}\ \bibnamefont
  {Castro~Neto}},\ }\bibfield  {title} {\bibinfo {title} {{Extrinsic Spin Hall
  Effect Induced by Resonant Skew Scattering in Graphene}},\ }\href
  {https://doi.org/10.1103/PhysRevLett.112.066601} {\bibfield  {journal}
  {\bibinfo  {journal} {Phys. Rev. Lett.}\ }\textbf {\bibinfo {volume} {112}},\
  \bibinfo {pages} {066601} (\bibinfo {year} {2014})}\BibitemShut {NoStop}%
\bibitem [{\citenamefont {Hsieh}\ \emph {et~al.}(2009)\citenamefont {Hsieh},
  \citenamefont {Xia}, \citenamefont {Qian}, \citenamefont {Wray},
  \citenamefont {Dil}, \citenamefont {Meier}, \citenamefont {Osterwalder},
  \citenamefont {Patthey}, \citenamefont {Checkelsky}, \citenamefont {Ong},
  \citenamefont {Fedorov}, \citenamefont {Lin}, \citenamefont {Bansil},
  \citenamefont {Grauer}, \citenamefont {Hor}, \citenamefont {Cava},\ and\
  \citenamefont {Hasan}}]{Hsieh_2009}%
  \BibitemOpen
  \bibfield  {author} {\bibinfo {author} {\bibfnamefont {D.}~\bibnamefont
  {Hsieh}}, \bibinfo {author} {\bibfnamefont {Y.}~\bibnamefont {Xia}}, \bibinfo
  {author} {\bibfnamefont {D.}~\bibnamefont {Qian}}, \bibinfo {author}
  {\bibfnamefont {L.}~\bibnamefont {Wray}}, \bibinfo {author} {\bibfnamefont
  {J.~H.}\ \bibnamefont {Dil}}, \bibinfo {author} {\bibfnamefont
  {F.}~\bibnamefont {Meier}}, \bibinfo {author} {\bibfnamefont
  {J.}~\bibnamefont {Osterwalder}}, \bibinfo {author} {\bibfnamefont
  {L.}~\bibnamefont {Patthey}}, \bibinfo {author} {\bibfnamefont {J.~G.}\
  \bibnamefont {Checkelsky}}, \bibinfo {author} {\bibfnamefont {N.~P.}\
  \bibnamefont {Ong}}, \bibinfo {author} {\bibfnamefont {A.~V.}\ \bibnamefont
  {Fedorov}}, \bibinfo {author} {\bibfnamefont {H.}~\bibnamefont {Lin}},
  \bibinfo {author} {\bibfnamefont {A.}~\bibnamefont {Bansil}}, \bibinfo
  {author} {\bibfnamefont {D.}~\bibnamefont {Grauer}}, \bibinfo {author}
  {\bibfnamefont {Y.~S.}\ \bibnamefont {Hor}}, \bibinfo {author} {\bibfnamefont
  {R.~J.}\ \bibnamefont {Cava}},\ and\ \bibinfo {author} {\bibfnamefont
  {M.~Z.}\ \bibnamefont {Hasan}},\ }\bibfield  {title} {\bibinfo {title} {{A
  tunable topological insulator in the spin helical Dirac transport regime}},\
  }\href {https://doi.org/10.1038/nature08234} {\bibfield  {journal} {\bibinfo
  {journal} {Nature}\ }\textbf {\bibinfo {volume} {460}},\ \bibinfo {pages}
  {1101} (\bibinfo {year} {2009})}\BibitemShut {NoStop}%
\bibitem [{\citenamefont {Araki}\ and\ \citenamefont
  {Nomura}(2017)}]{araki_2017}%
  \BibitemOpen
  \bibfield  {author} {\bibinfo {author} {\bibfnamefont {Y.}~\bibnamefont
  {Araki}}\ and\ \bibinfo {author} {\bibfnamefont {K.}~\bibnamefont {Nomura}},\
  }\bibfield  {title} {\bibinfo {title} {{Skyrmion-induced anomalous Hall
  conductivity on topological insulator surfaces}},\ }\href
  {https://doi.org/10.1103/PhysRevB.96.165303} {\bibfield  {journal} {\bibinfo
  {journal} {Phys. Rev. B}\ }\textbf {\bibinfo {volume} {96}},\ \bibinfo
  {pages} {165303} (\bibinfo {year} {2017})}\BibitemShut {NoStop}%
\bibitem [{\citenamefont {Ishizuka}\ and\ \citenamefont
  {Nagaosa}(2018)}]{ishizuka_2018}%
  \BibitemOpen
  \bibfield  {author} {\bibinfo {author} {\bibfnamefont {H.}~\bibnamefont
  {Ishizuka}}\ and\ \bibinfo {author} {\bibfnamefont {N.}~\bibnamefont
  {Nagaosa}},\ }\bibfield  {title} {\bibinfo {title} {{Spin chirality induced
  skew scattering and anomalous Hall effect in chiral magnets}},\ }\href
  {https://doi.org/10.1126/sciadv.aap9962} {\bibfield  {journal} {\bibinfo
  {journal} {Science Advances}\ }\textbf {\bibinfo {volume} {4}},\ \bibinfo
  {pages} {eaap9962} (\bibinfo {year} {2018})}\BibitemShut {NoStop}%
\bibitem [{\citenamefont {Tokura}\ and\ \citenamefont
  {Nagaosa}(2018)}]{Tokura_2018}%
  \BibitemOpen
  \bibfield  {author} {\bibinfo {author} {\bibfnamefont {Y.}~\bibnamefont
  {Tokura}}\ and\ \bibinfo {author} {\bibfnamefont {N.}~\bibnamefont
  {Nagaosa}},\ }\bibfield  {title} {\bibinfo {title} {{Nonreciprocal responses
  from non-centrosymmetric quantum materials}},\ }\href
  {https://doi.org/10.1038/s41467-018-05759-4} {\bibfield  {journal} {\bibinfo
  {journal} {Nature Communications}\ }\textbf {\bibinfo {volume} {9}},\
  \bibinfo {pages} {3740} (\bibinfo {year} {2018})}\BibitemShut {NoStop}%
\bibitem [{\citenamefont {Dzyaloshinsky}(1958)}]{DZYALOSHINSKY_1958}%
  \BibitemOpen
  \bibfield  {author} {\bibinfo {author} {\bibfnamefont {I.}~\bibnamefont
  {Dzyaloshinsky}},\ }\bibfield  {title} {\bibinfo {title} {{A thermodynamic
  theory of “weak” ferromagnetism of antiferromagnetics}},\ }\href
  {https://doi.org/https://doi.org/10.1016/0022-3697(58)90076-3} {\bibfield
  {journal} {\bibinfo  {journal} {Journal of Physics and Chemistry of Solids}\
  }\textbf {\bibinfo {volume} {4}},\ \bibinfo {pages} {241} (\bibinfo {year}
  {1958})}\BibitemShut {NoStop}%
\bibitem [{\citenamefont {Moriya}(1960)}]{moriya_1960}%
  \BibitemOpen
  \bibfield  {author} {\bibinfo {author} {\bibfnamefont {T.}~\bibnamefont
  {Moriya}},\ }\bibfield  {title} {\bibinfo {title} {{Anisotropic Superexchange
  Interaction and Weak Ferromagnetism}},\ }\href
  {https://doi.org/10.1103/PhysRev.120.91} {\bibfield  {journal} {\bibinfo
  {journal} {Phys. Rev.}\ }\textbf {\bibinfo {volume} {120}},\ \bibinfo {pages}
  {91} (\bibinfo {year} {1960})}\BibitemShut {NoStop}%
\end{thebibliography}%

\end{document}